\documentclass[preprint2]{proto}
\usepackage{times}

\newcommand{\refs}{\par\noindent\hangindent=1pc\hangafter=1}
\def\epstwover@scaling{0.3}

\def\plottwover#1#2{
  \centering
  \leavevmode
  \includegraphics[height={\epstwover@scaling\textheight}]{#1} \hfil \,
  \includegraphics[height={\epstwover@scaling\textheight}]{#2} \hfil \,
}

\begin{document}

\title{\textbf{\LARGE Disk Winds, Jets, and Outflows: 
   Theoretical and Computational Foundations}}

\author {\textbf{\large Ralph E. Pudritz}}
\affil{\small\em McMaster University}
\author {\textbf{\large Rachid Ouyed}}
\affil{\small\em University of Calgary}
\author {\textbf{\large Christian Fendt}}
\affil{\small\em Max-Planck Institute for Astronomy, Heidelberg }
\author {\textbf{\large Axel Brandenburg}}
\affil{\small\em Nordic Institute for Theoretical Physics}

\begin{abstract}
\baselineskip = 11pt
\leftskip=0.65in
\rightskip=0.65in
\parindent=1pc
{\small 
We review advances in the theoretical and
computational studies of disk winds, jets
and outflows including:
the connection between accretion and jets, the launch of jets from
magnetized disks, the coupled evolution
of jets and disks,
the interaction of magnetized young stellar objects
with their surrounding disks and the relevance to outflows,
and finally, the link between jet formation and gravitational collapse.
We also address the predictions that the theory 
makes about jet kinematics, collimation, and rotation, 
that have recently been confirmed by 
high spatial and spectral resolution
observations.
Disk winds have a universal character 
that may account for jets and outflows 
during the formation of massive stars as well as brown dwarfs.
 \\~\\~\\~}
 
\end{abstract}  

\section{\textbf{INTRODUCTION}}

The close association of jets and outflows with 
protostellar accretion disks 
is one of the hallmarks   
of the accretion picture of low mass star formation.
The most energetic outflow phase occurs 
during gravitational collapse of a molecular 
cloud core - the 
so-called Class 0 phase - when much of its envelope 
is still raining down onto the forming protostellar disk 
and the disk accretion rate is high.
Later, in the T-Tauri star (TTS) stage, when most of the original core has 
been accreted and the young
stellar object (YSO) is being fed by lower accretion rates through 
the surrounding Keplerian accretion disk, the   
high-speed jet becomes  
optically visible. When the disk disappears in the 
weak-lined TTS (WTTS) phase,
the jet goes with it.  

The most comprehensive theoretical picture that 
we have for these phenomena is that jets are 
highly collimated, hydromagnetic disk winds whose 
torques efficiently extract disk angular momentum and 
gravitational potential energy.
Jets also sweep up ambient
molecular gas and drive large scale 
molecular outflows.     
A disk wind was first suggested as the origin of jets 
from accretion 
disks around black holes 
in the seminal 
paper by {\it Blandford and Payne} 
(1982, BP82), and was soon proposed as the mechanism for 
protostellar jets 
({\it Pudritz and Norman}, 1983, 1986).

Several major observational breakthroughs have taken place 
in the study of jets and outflows 
since PPIV (held in 1998). 
Direct, high resolution spectro-imaging and 
adaptive optics methods discovered   
the rotation of protostellar jets 
({\it Bacciotti et al.}, 2003).  
These observations also revealed that jets
have an onion-like, velocity structure (with 
the highest speeds being closest to the outflow axis).   
This work provides strong support for the    
idea that jets originate as centrifugally driven
MHD winds from extended regions of their surrounding disks 
(see chapter by {\it Ray et al.}).
Recently, outflows and disks have also been discovered around 
massive stars (see chapter by {\it Arce et al.}) as well as brown dwarfs
(e.g., {\it Bourke et al.} 2005) 
implying that the mechanism is important across the entire
stellar mass spectrum.

Major advances in the theoretical modeling of these systems
have also occurred, due primarily to a variety   
of MHD computational studies.  Simulations now
resolve the global evolution of disks and outflows, track
the interaction of disks with central
magnetized stars, and even follow the generation of outflows
during gravitational collapse.  These studies    
shows that  
jets and disks are closely coupled  
and that jet dynamics  
scales to YSOs of all masses.   

Our review examines the theory of the central engine
of jets and its exploration through the use of 
computer simulations.  We 
focus mainly on developments since PPIV and refer to the 
review by {\it K\"onigl and Pudritz} (2000, KP00) 
for a discussion of  
the earlier literature, as well as {\it Pudritz} (2003) and 
{\it Heyvaerts} (2003) for more technical background.
We first   
discuss the basic theory of disk winds and their
kinematics.  
We then switch to   
computational studies of
jets from accretion disks treated as  
boundary conditions, as well as global
simulations including the disk.
We then examine the 
innermost regions of the disk where 
the stellar magnetosphere 
interacts with the disk, as well as
the surface of the star that may drive an accretion-powered
outflow.  
Finally, we 
discuss how outflows are generated
during the early stages of the gravitational collapse. 

\section{\textbf{THEORY OF DISK WINDS}} 

An important insight into the 
nature of the engine for jets can be gleaned from the observed ratio of  
the momentum transport rate (or thrust) carried by the CO 
molecular outflow to the    
thrust that can be provided by the bolometric luminosity of the central star  
(e.g., {\it Cabrit and Bertout}, 1992);
\begin{equation}
F_{outflow}/F_{rad} = 250 (L_{\rm bol}/10^3 L_{\odot})^{-0.3}, 
\end{equation}
This relation 
has been confirmed and extended by the analysis of data
from over 390 outflows, ranging over six decades
up to $10^6L_{\odot}$ in
stellar luminosity ({\it Wu et al.}, 2004).
It suggests that jets from both low and high mass systems are probably 
driven by a single, non-radiative,  
mechanism.  

Jets are observed to have 
a variety of structures and time-dependent behaviour -- from
internal shocks
and moving knots to systems of bow shocks that suggest
long-time episodic outbursts.
They show wiggles and often have cork-screw like structure, 
suggesting the presence either of jet precession, 
the operation of non-axisymmetric kink modes, or both.  
Given the highly nonlinear behaviour of
the force balance equation for jets (the so-called Grad-Shafranov
equation), theoretical work has focused on tractable 
and idealized time-independent, and axisymmetric or 
self-similar models
(e.g., BP82) of various kinds.  

\subsection{\textbf{Conservation Laws and Jet Kinematics }} 

Conservation laws are the gold standard in physics,
and play a significant role in understanding astrophysical jets.
This is because whatever the details (e.g.\ the asymptotics,
the crossing of critical points, the way that matter 
is loaded onto field lines within the disks, etc.), conservation laws
strongly constrain the flux of mass, angular momentum, and
energy. 
What cannot be constrained by these laws will depend on 
the general physics of the disks such as on  
how matter is loaded onto field lines. 

Jet dynamics can be described by the time-dependent,
equations of ideal MHD.  The evolution of a magnetized,
rotating system that is threaded by a large-scale field ${\bf B}$
involves (i) the 
continuity equation for  
a conducting gas of
density $\rho$ moving at velocity
${\bf v} $ (which includes turbulence);  
(ii) the equation of motion for the gas
which undergoes 
pressure ($p$), gravitational 
(with potential $\Phi$),
and Lorentz forces; 
(iii) the induction equation
for the evolution of the magnetic field in the moving
gas where the current density is ${\bf j} = (c / 4 \pi) {\bf \nabla \times B}$; 
(iv) the energy equation, 
where $e$ is the internal energy per unit mass;  
and, (v) the absence of magnetic monopoles. These are   
written as: 
\begin{eqnarray}
{\partial\rho\over\partial t}+\nabla .(\rho{\bf v}) &=&0 \qquad  \\
\rho \left({\partial{\bf v}\over \partial t}+({\bf v.\nabla}){\bf v}\right)
+\nabla p +\rho {\bf \nabla}\Phi -
{{\bf j}\times {\bf B}\over c}&=&0 \qquad \\
{\partial {\bf B}\over\partial t}-\nabla\times ({\bf v}\times {\bf B})&=&0 \qquad  \\
\rho\left({\partial{ e}\over \partial t}+({\bf v.\nabla)}e\right)
+ p({\bf \nabla .v})&=&0 \qquad \\
\nabla .{\bf B}&=&0 \qquad
\end{eqnarray}

We specialize to the restricted case of  
stationary, as well as  2D (axisymmetric) flows, from 
which the conservation laws follow. 
It is useful to decompose vector quantities into poloidal
and toroidal components (e.g.\ magnetic field 
$  {\bf B =  B_p} +  B_{\phi}
{\bf \hat e_{\phi}} $).  In axisymmetric conditions, the 
poloidal field ${\bf B_p}$ can be derived from a single scalar potential
$a(r,z)$ whose individual values, $a=\mbox{const}$, define the 
surfaces of constant magnetic flux in the outflow and can be
specified at the surface of the disk 
(e.g., {\it Pelletier and Pudritz}, 1992; PP92).   

{\bf Conservation of mass and magnetic flux} along a field line
can be combined into a single function
$k$ that is called the ``mass load'' of the wind
which is a constant along a magnetic field line; 
\begin{equation}
\rho {\bf v_p} = k {\bf B_p}.
\end{equation}
\noindent
This function represents the mass load per unit time,
per unit magnetic flux
of the wind. For axisymmetric flows, 
its value is preserved on each ring of  
field lines emanating from the accretion disk. 
Its value on each field line is determined
by physical conditions - including dissipative
processes - near the disk surface.
It may be more revealingly written as
\begin{equation}
k(a) = { \rho v_p \over B_p} = {d \dot M_{\rm w} \over d \Psi},
\end{equation}
where $d\dot M_{\rm w}$ is the mass flow rate through an annulus
of cross-sectional area $dA$ through the wind and $d\Psi$
is the amount of poloidal magnetic flux threading through this same annulus.
The mass load profile is a function
of the footpoint radius $r_0$ of the wind on the disk. 

The toroidal field in rotating flows derives from the induction
equation;
\begin{equation}
B_{\phi} = {\rho  \over k} ( v_{\phi} - \Omega_0 r),
\end{equation}
where
$ \Omega_0$ is the angular
velocity of the disk at the mid-plane.
This result
shows that the strength of the toroidal field in the jet
depends on the mass loading as well as the jet density.
Denser winds should have stronger toroidal fields.
We note however, that the density does itself depend on the value 
of $k$.  Equation (9) also suggests that at 
higher mass loads, one has lower toroidal field strengths. 
This can be reconciled however, since it can be shown from the conservation laws
(see below) that the value of $k$ is related to the density of the outflow
at the Alfv\'en point on a field line; $k = (\rho_A/4\pi)^{1/2}$
(e.g., PP92).
Thus, higher mass loads correspond to denser winds and when
this is substituted into equation (9), we see that this also implies 
stronger toroidal fields.  
We show later that jet collimation depends on hoop stress
through the toroidal field and thus 
the mass load must have a very important effect on
jet collimation 
(Section 3.2). 

{\bf Conservation of angular momentum} along each
field line leads to the conserved angular momentum per unit mass;
\begin{equation}
l(a) = r v_{\phi} - {r B_{\phi} \over 4 \pi k} = \mbox{const}. 
\end{equation}
The form for $l$ reveals that the total angular momentum
is carried by both the rotating gas (first term) as well
by the twisted field (second term), the relative proportion
being determined by the mass load.

The value of $l(a)$ that is  
transported along each field line is fixed by the
position of the Alfv\'en point in the flow, where the
poloidal flow speed reaches the Alfv\'en speed for the
first time
($m_{\rm A}=1$). It is easy to show that
the value of the specific angular momentum is;
\begin{equation}
l(a) = \Omega_0 r_{\rm A}^2 = (r_{\rm A}/r_0)^2 l_0.
\end{equation}
where 
$l_0 = v_{K,0}r_0 = \Omega_0 r_0^2$ is the 
specific angular momentum of a Keplerian disk.
For a field line starting at a point $r_0$ on the
rotor (disk in our case), the Alfv\'en radius is
$r_{\rm A}(r_0)$ and constitutes a lever arm for the flow.
The result shows that the angular momentum per unit
mass that is being extracted from the
disk by the outflow is a factor of $(r_{\rm A}/r_0)^2$ greater than
it is for gas in the disk.  For typical 
lever arms, one particle in the outflow can
carry the angular momentum of ten of its fellows left behind
in the disk.

{\bf Conservation of energy} along a field line
is expressed as a generalized version of Bernoulli's theorem
(this may be derived by taking the dot product of the
equation of motion with ${\bf B_p} $).  
Thus, there is a specific
energy $e(a)$ that is a constant along field lines, which
may be found in many papers (e.g., \ BP82 and PP92). 
Since the terminal speed $v_p = v_{\infty}$ of the disk
wind is much greater than its rotational speed, and
for cold flows, the pressure may also
be ignored, one finds the result: 
\begin{equation}
v_{\infty} \simeq 2^{1/2} \Omega_0 r_{\rm A} = (r_{\rm A}/r_0) v_{\rm esc,0}. 
\end{equation}

There are three important consequences for jet kinematics here;  
(i) that the terminal speed exceeds the {\it }local escape speed from 
its launch point on the disk by the lever arm ratio; 
(ii) the terminal speed scales with the Kepler speed as a function 
of radius, so that the flow will have an onion-like layering of
velocities, the largest inside, and the smallest on the larger scales,
as seen in the observations; and
(iii) that the terminal speed depends on the depth
of the local gravitational well at the footpoint of the
flow -- implying that it is essentially scalable to flows from disks around
YSOs of any mass and therefore universal.  

Another useful form of the conservation laws is the combination
of energy and angular momentum conservation to produce a new
constant along a field line (e.g., \ PP92);
$j(a) \equiv e(a) - \Omega_0  l(a) $.     
This expression has been used ({\it Anderson et al.}, 2003) to 
deduce the rotation rate of the launch region on the Kepler
disk, where the observed jet rotation speed is $v_{\phi, \infty}$
at a radius $r_{\infty}$ and which is moving in the poloidal direction with 
a jet speed of $v_{p, \infty}$. 
Evaluating $j$ for a cold jet at infinity
and noting that its value (calculated at the foot point)
is $j(a_0)=-(3/2) v_{\rm K,0}^2$, one solves for the Kepler rotation at the 
point on the disk where this flow was launched: 
\begin{equation}
\Omega_0 \simeq v_{p,\infty}^2/ \left(2 v_{\phi,\infty} r_{\infty}\right). 
\end{equation}
When applied to the observed rotation of the 
Large Velocity Component (LVC) of the jet   
DG Tau ({\it Bacciotti et al.}, 2002), this yields a range of disk radii 
for the observed rotating material 
in the range of disk radii, 0.3--4 AU, and the magnetic  
lever arm is $r_{\rm A}/r_0 \simeq 1.8$--$2.6$.

\subsection{\textbf{Angular Momentum Extraction}}

How much angular momentum can such a wind extract from the disk? 
The angular momentum equation for the accretion disk undergoing
an external magnetic torque may be written:

\begin{equation}
\dot M_{\rm a} { d (r_0 v_0) \over dr_0} = - r_0^2 B_{\phi} B_z \vert_{r_0, H},
\end{equation}
\noindent
where we have ignored transport by MRI turbulence or spiral waves.
By using the relation between poloidal field and outflow on the 
one hand, as well as the link between the toroidal field and 
rotation of the disk on the other, the angular momentum equation 
for the disk yields one of the most profound scaling relations in disk wind
theory -- namely -- the link between disk accretion and mass outflow
rate (see KP00, PP92 for details):
\begin{equation}
\dot M_{\rm a} \simeq (r_{\rm A}/ r_0)^2 \dot M_{\rm w}.
\end{equation}
The observationally well known result that in many systems,
$\dot M_{\rm w} / \dot M_{\rm a} \simeq 0.1$ is a consequence of the 
fact that lever arms are often found in numerical and
theoretical work to be $r_{\rm A}/r_0 \simeq 3$ -- the observations
of DG Tau being a perfect example.
Finally, we note that the angular momentum that is observed to be carried by 
these rotating flows (e.g.\ DG Tau) is a consistent fraction
of the excess disk angular momentum -- from 60--100\%
(e.g., {\it Bacciotti}, 2004), which is consistent with the high
extraction efficiency discussed here.

\subsection{\textbf{Jet Power and Universality }}

These results can be directly connected to 
the observations of momentum and energy transport in the molecular
outflows.
Consider the total mechanical power that is carried
by the jet, which may be written as (e.g., {\it Pudritz}, 2003);
\begin{equation}
L_{\rm jet}= {\textstyle{1 \over 2}}
\int_{r_{\rm i}}^{r_{\rm j}} d\dot M_{\rm w} v_{\infty}^2
\simeq {G M_* \dot M_{\rm a} \over 2r_{\rm i}}
\left(1 - {r_{\rm i}^2 \over r_{\rm j}^2}\right)
\simeq {\textstyle{1 \over 2}} L_{\rm acc} .
\end{equation}
This explains the observations of Class 0 outflows
wherein $L_{\rm w}/L_{\rm bol} \simeq 1/2$, 
since the main luminosity of the central source at this time 
is due to accretion and not nuclear reactions.
(The factor of $1/2$ arises from the dissipation
of some accretion energy as heat at the inner boundary).
The ratio of wind to stellar luminosity decreases at later 
stages because the accretion luminosity
becomes 
relatively small compared to the bolometric luminosity of the 
star as it nears the ZAMS. 

This result states that the wind luminosity taps the gravitational
energy release through accretion in the gravitational potential of
the central object -- and is a direct consequence of Bernoulli's
theorem.  This, and the previous results, imply 
that jets may be produced in any 
accreting system.  The lowest mass outflow that has yet
been observed corresponds to a proto-brown dwarf of
luminosity $\simeq 0.09 L_{\odot}$, a stellar 
mass of only $ 20 - 45 M_{Jup}$, 
and a very low mass disk $ < 10^{-4} M_{\odot}$ ({\it Bourke et al.}, 2005).

It should be possible, therefore, for lower luminosity jets to be    
launched from the disks
around Jovian mass planets. 
Recent hydrodynamical simulations of circumstellar accretion disks        
containing and building up an orbiting protoplanetary core have
numerically proven the existence of a circum-planetary sub-disk in        
almost Keplerian rotation close to the planet ({\it Kley et al.}, 2001).
The accretion rate of these sub-disks is about                            
$\dot{M}_{\rm cp} = 6\times 10^{-5}\,M_{Jup}\,{\rm yr}^{-1}$
and is confirmed by many independent simulations.                         
With that, the circum-planetary disk temperature may reach values         
up to $2000\,$K indicating a sufficient degree of ionization for          
matter-field coupling and would also allow for strong equipartition       
field strength ({\it Fendt}, 2003).                                       

The possibility of a planetary scale MHD outflow, similar to 
the larger scale YSO disk winds, is indeed quite likely because:
(i) the numerically established existence of circum-planetary disks
is a natural feature of the formation of massive planets;
and (ii) the feasibility of a large-scale magnetic field in the               
  protoplanetary environment                                                
({\it Quillen and Trilling}, 1998; {\it Fendt}, 2003).    
One may show, moreover, that 
the outflow velocity is of the order of the escape speed for the  
 protoplanet, at about                                                     
$60\,{\rm km\,s}^{-1}$ ({\em Fendt}, 2003).                           

On very general grounds, disk winds are also likely 
to be active during massive star formation
(e.g., {\it K\"onigl} 1999).  Such outflows may already 
start during the early collapse phase when the central YSO 
still has only a fraction 
of a solar mass (e.g., \ {\it Pudritz
and Banerjee}, 2005).  Such early outflows may
actually enhance the formation of massive stars via 
disk accretion by punching a hole in the infalling 
envelop and releasing the building radiation pressure
(e.g., \ {\it Krumholz et al.}, 2005).

\subsection{\textbf{Jet Collimation}}  

In the standard picture of hydromagnetic winds,
collimation of an outflow occurs
because of the increasing
toroidal magnetic field in the flow resulting from 
the inertia of the gas. 
Beyond the Alfv\'en surface, 
equation~(8) shows that the ratio of the 
toroidal field to the poloidal
field in the jet 
is of the order
$B_{\phi} / B_p \simeq r/ r_{\rm A} \gg 1$, so that the field becomes
highly toroidal.
In this situation, collimation is achieved
by the tension force associated with the toroidal field
which leads to a radially inwards directed component
of the Lorentz force (or ``$z$-pinch");
$ F_{\rm Lorentz, r} \simeq j_z B_{\phi}$.
The stability of such systems is examined in the next
section.

In {\it Heyvaerts and Norman} (1989) it was shown
that two types of solution are possible depending upon
the asymptotic behaviour of the total current intensity in the jet;
\begin{equation}
 I = 
 2 \pi \int_0^r j_z(r',z')dr' = (c/2)
 r B_{\phi}. 
\end{equation}
In the limit that $I \rightarrow 0$ as
$r \rightarrow \infty $, the field lines are paraboloids
which fill space.  On the other hand, if the current
is finite in this limit, then the flow is collimated to cylinders.
The collimation of a jet therefore depends upon
its current distribution -- and hence on the radial distribution
of its toroidal field -- which, as we saw earlier,
depends on the mass load.  Mass loading therefore must play
a very important role in controlling jet collimation. 

It can be shown ({\it Pudritz et al.}, 2006; PRO)
that for a power law distribution of the magnetic field
in the disk,  
$B_z(r_0, 0) \propto r_0^{\mu - 1}$ and an 
injection speed at the base of a (polytropic) corona that scales as 
the Kepler speed,
that the mass load takes the form
$k \propto r_o^{-1-\mu}$.
In this regime, 
the current takes the form
$ I(r,z) \propto r_o^{-\mu - (1/2)}$.
Thus, the current goes to zero for models with $\mu < -1/2$,
and that these therefore must be wide angle flows.  For models
with $\mu > -1/2$ however, the current diverges, and the 
flow should collimate to cylinders. 

These results predict that jets
should show different degrees of collimation 
depending on how they are mass loaded (PRO).  As an example,
neither the highly centrally concentrated, magnetic field lines 
associated with the initial split-monopole magnetic 
configuration used in simulations by {\it Romanova et al.}\ (1997), nor 
the similar field structure invoked
in the X-wind (see review by {\it Shu et al.}, 2000) should become
collimated in this picture.  
On the other hand, less centrally (radially) concentrated
magnetic configurations such as the potential configuration of
{\it Ouyed and Pudritz} (1997a, OPI) and BP82 should collimate to cylinders. 

This result also explains the range of collimation that is 
observed for molecular outflows.  
Models for observed outflows fall into two general categories: the
jet-driven bow shock picture, and a wind-driven 
shell picture in which the molecular gas is driven by  
an underlying wide-angle wind component such as given
by the X-wind 
(see review by {\it Cabrit et al.}, 1997). 
A survey of molecular outflows by {\it Lee et al.}\ (2000) found
that both mechanisms are needed in order to explain the full
set of systems observed. 

Finally, we note that apart from these general theorems on collimation,
{\it Spruit et al.}\ (1997) has proposed that a sufficiently strong poloidal
field that is external to the flow could also force its collimation.
According to this criterion, such a poloidal
field could collimate jets 
provided that the field strength decreases no slower than
$B_{\rm p} \sim r^{-\mu}$ with $\mu\le1.3$.

\section{\textbf{SIMULATIONS: DISKS AS JET ENGINES}}

Computational approaches are necessary if we are to 
open up and explore the vast spaces of solutions to    
the highly nonlinear jet problem.
The first simulations of nonsteady MHD jets from
accretion disks were performed by {\it Uchida and Shibata} (1985)
and {\it Shibata and Uchida} (1986). 
These early simulations were based on 
initial states that were violently out of equilibrium
and demonstrated the role of magnetic fields in  
launching and accelerating jets to velocities of order of the Keplerian 
velocity of the disk.

The published simulations of non-radiative ideal MHD YSO jets 
differ in their assumed initial conditions, 
such as the magnetic field distribution on the disk, 
the conditions in the  plasma  above the disk surfaces, 
the state of the initial disk corona, and the handling of
the gravity of the central star. Nevertheless, they share common goals: 
to establish and verify the four important stages in jet 
evolution, namely,  (i) ejection; (ii) acceleration; 
(iii) collimation; and (iv) stability.  
In the following, we describe a basic physical approach
for setting up 
clean numerical simulations and how this leads to  
advances in our understanding of disk winds and outflows.
 
\subsection{2-Dimensional Simulations }

 The simulations we  discuss here have been studied in greater 
detail in {\it Ouyed and Pudritz} (1997a; OPI); 
{\it Ouyed and Pudritz} (1997b; OPII); 
{\it Ouyed et al.}\ (1997; OPS); {\it Ouyed and Pudritz} (1999; OPIII);   
and PRO. 
We use these for  
pedagogical purposes in this review.
To see animations of the simulations presented here 
as well as those performed by other authors,  
the interested reader is directed to the 
``animations" link at ``{\it http://www.capca.ucalgary.ca}" 
(hereafter CAPCA). These simulations were  run using the ZEUS 
(2-D, 3-D and MP) code   which is arguably the best documented and 
utilized MHD code in the literature 
({\it Stone and Norman}, 1992, 1994).  
It is an explicit, finite difference code that runs on a staggered grid.  
The equations generally solved are those of ideal MHD listed in Section 1,
equations~(2)--(6).

The evolution of the magnetic field (induction equation~(4) above)  
is followed by the method of constrained transport.  
In this approach, if ${\bf \nabla . B} = 0$ holds for 
the initial magnetic configuration, then it remains so for all 
later times to machine accuracy.  The obvious way of securing this 
condition is to use an initial vector potential ${\bf A(r,z},t=0)$  
that describes the desired initial magnetic field at every point in the 
computational domain.

To ensure a stable initial state 
that allows for a tractable simulation,  
three simple rules are useful 
(OPS and OPI): (i) use a corona that in hydrostatic balance 
with the central object (to ensure a perfectly stable 
hydrostatic equilibrium of the corona, the point-mass
gravitational potential should be relocated to zone centers
- see Section 4.2 in OPI for details);  
(ii) put the disk in pressure balance with the corona above it, and 
(iii) use a force-free magnetic field configuration 
to thread the disk and corona.  The initial magnetic 
configurations should be chosen so that no 
Lorentz force is exerted on the initial  (non-rotating) 
hydrostatic corona described above.

The BP82 self-similar solution for jets uses 
a simple polytropic equation of state, with an index
$\gamma = 5/3$.  Their solution is possible because 
this choice eliminates the many complications that arise from the
the energy equation, while preserving the essence of the 
outflow problem. 
This Ansatz corresponds 
to situations where the combined effects of  heating and cooling 
simulates a tendency toward a locally constant entropy.
This choice 
also greatly simplifies the numerical setup
and allows one to test the code against analytic solutions.  

Another  key simplification in this approach 
is to examine the physics of the outflow for
fixed physical conditions in the accretion disk. 
Thus, the accretion disk at the base of 
the corona -- and in pressure balance with the overlying
 atmosphere -- is given a density profile
that it maintains at all times in the simulation, since the disk
boundary conditions are applied to the ``ghost zones"
and are not part of the computational domain. 
In part, this simplification may be justified by the 
fact that typically, accretion disks will evolve on longer time scales than 
their associated jets.    

The hydrostatic state one arrives at has a simple analytic solution
which was used as the initial state for all of our simulations and
was adopted and further developed by several groups.
These studies include e.g. low plasma-$\beta$ monopole-like field
distributions ({\it Romanova et al.}, 1997);
a magneto-gravitational switch mechanism ({\it Meier et al.}, 1997);
the disk accretion-ejection process ({\it Kudoh et al.}, 1998);
the interrelation between the grid shape and the degree of
flow collimation ({\it Ustyugova et al.}, 1998);
a self-adjusting disk magnetic field inclination ({\it Krasnopolsky et
al.}, 1999, 2003);
the interrelation between the jet's turbulent magnetic diffusivity and
collimation
({\it Fendt and {\v C}emelji{\'c}}, 2002);
a varation of the disk rotation profile ({\it Vitorino et al.}, 2002);
or dynamo maintained disk magnetic fields ({\it von Rekowski and
Brandenburg}, 2004).
More elaborate setups wherein the initial magnetic field 
configuration originates on the 
surface of a star and connects with the disk 
have also been examined (Section 5 -- e.g.,   
{\it Hayashi et al.}, 1996; {\it Goodson et al.}, 1997; 
{\it Fendt and Elstner}, 1999; {\it Keppens and Goedbloed}, 2000).   

The lesson from these varied simulations seems to be that all
roads lead to a disk field.  Thus,
the twisting of a closed magnetic 
field that initially threads the disk beyond the 
co-rotation radius rapidly inflates the field and then  
disconnects it from the star, thereby 
producing an open  
disk-field line. Likewise, simulations including dynamo 
generated fields in the disk lead to a state that 
resembles our initial setup (see Section 4).  
Thus, from the numerical point 
of view,  a ``fixed-disk" simulation is 
general and useful.

\begin{figure}[t!]
\epsscale{1.0}
\begin{center}
\includegraphics[width=\columnwidth]{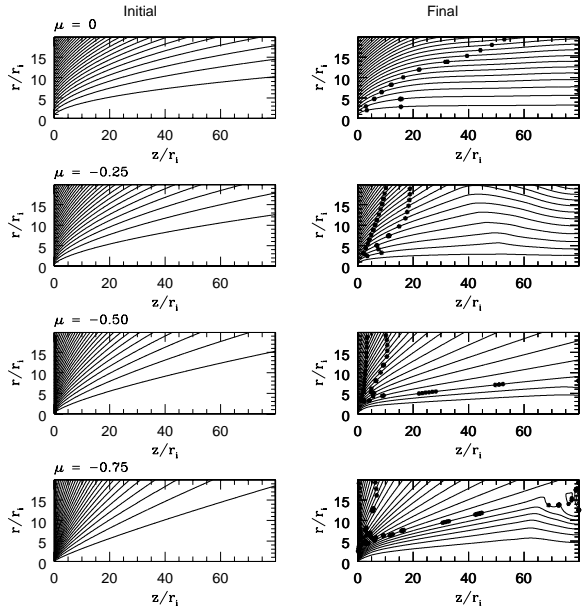}
\end{center}
\caption{\small Left panels: initial magnetic field configurations
         for winds with $\mu = 0$ (OPI), $\mu = -0.25$ (BP),  
	 $\mu = -0.5$ (PP),
          and $\mu = -0.75$ (steep) 
Right panels:
       final magnetic field configurations (at $t = 400$)  
         for each case, with Alfv\'en
          points (filled circles) and fast magneto-sonic points (stars) marked.
          Note the more open magnetic -- and  
            streamline -- structures as $\mu$ goes down. [Adapted
	    from PRO].}
\label{fig:diff_B}
\end{figure}

The setup requires  five physical  
quantities to be specified at all points of the disk surface 
at all times (see OPI for all details).  
These are the disk density $\rho(r_0)$; 
components of the vertical and toroidal magnetic field, 
$B_z(r_0)$ and $B_{\phi}(r_0)$; and velocity components in the 
disk, $v_z(r_0)$ and $v_{\phi}(r_0)$, 
where $r_0$ is the radius (cylindrical coordinates
are adopted with the disk located at 
$z=0$; $r_0=r(z=0)$).  The remaining field
component $B_r(r_0)$ is determined by the solenoidal 
condition, while the radial inflow speed through the disk
is neglected ($v_r(r_0) \simeq 0$) since
it is far smaller than the sound speed in a real disk. 
The model is described by five parameters 
defined at the inner disk radius $r_{\rm i}$, three of which describe
the initial corona.  
The two additional parameters describe the disk physics and  
are $\nu_{\rm i}$, 
which scales the toroidal field in the disk 
$B_{\phi} = \nu_{\rm i}\times  (r_{\rm i}/r_0)$, 
and the (subsonic) injection speed of material
from the disk into the base of the corona, 
$v_{\rm inj}= v_z(r_0)/v_{\phi}(r_0) \simeq 0.001$. 
Simulations were  
typically run with $(500 \times 200)$ 
spatial zones  to resolve  a physical region of 
$ (80 r_{\rm i} \times 20 r_{\rm i} ) $ in
the $z$ and $r$ directions, respectively. 
A  resolution of 
10 zones per $r_{\rm i}$ provides 
enough dynamical range to accurately follow
the smooth  acceleration 
above the disk surface. The simulations were run 
up to $400 t_{\rm i}$
(where $t_{\rm i}$ is the Kepler time for an orbit 
at the inner edge of the disk).

A series of magnetic configurations is shown in 
Fig.~\ref{fig:diff_B}
where the disk field is modeled as
$B_{z}(r_0,0)\propto r_0^{\mu -1}$.  
The initial configurations
range  from the rather well collimated 
(such as the potential configuration of OP97, 
$\mu =0.0$; and the Blandford- Payne  configuration, $\mu =-0.25$), to 
the initially more open configurations 
of Pelletier-Pudritz (PP92; $\mu=-0.5$), and steeper
($\mu=-0.75$).  

A good example of a configuration that evolves into 
a stationary jet was studied by 
OPI and is shown in the upper
panel of Fig.~\ref{fig:diff_B}).
The simulation shows the existence of an 
acceleration region very close to the disk surface.    
The acceleration from the disk occurs by a centrifugal 
effect whereby, at some point along sufficiently inclined field lines, 
centrifugal force dominates gravity and gas is 
flung away like beads on a wire. Thus, a toroidal field 
component is created because the field lines corotate 
with the underlying disk.   The inertia of the matter 
in the flow region ultimately forces the field to fall 
behind the rotation of the disk, which produces the 
toroidal field component.    
More precisely, beyond the Alfv\'en surface 
(shown as dots in the right hand panels in Fig.~\ref{fig:diff_B}), 
the  hoop stress induced by the self-generated 
$B_{\phi}$ eventually dominates, which provides the collimation. 
The ratio of the toroidal  to poloidal magnetic field  
along illustrative field lines (e.g., \ Figs.~3 in OPS) 
clearly shows that the predominant magnetic field in  
jets beyond their fast magnetosonic (FM) 
surfaces is the toroidal field component.   
The gas is eventually collimated into cylinders 
parallel to the disk's axis starting at the 
Alfv\'en surface to distances much  beyond the FM surface 
(see Fig.~\ref{fig:diff_B}). 
   
The final velocities achieved by such winds 
are of the order of 2 times the Kepler velocity, 
along a given field line (e.g.,  Fig.~3 in  OPS), 
which translates to roughly 100-300 km s$^{-1}$ 
for a standard YSO. The general trend is that the 
jet solutions, dominated mainly by the poloidal 
kinetic energy, are very efficient in magnetically extracting 
angular momentum and energy from the disk, as confirmed 
by simulations performed by other groups.
In 1000 years, for example, and for a standard YSO, 
the simulations imply that  the disk winds  can carry 
a total energy of $3 \times 10^{43}$ ergs, sufficient 
to produce the observed molecular outflows
 
In OPII an initial vertical field configuration, 
was used.  
The simulations show that the 
strong toroidal magnetic field generated 
recollimates the flow towards the disk's axis and, 
through MHD shocks, produces knots. 
The knot generator is located at a 
distance of about $z\simeq 8r_{\rm i}$ from the surface of the disk (OPS). 
Knots propagate down the length of the  jet at speeds 
less than the diffuse component of the outflow. 
The knots are episodic, 
and are intrinsic to the jet and not the accretion disk, in this calculation.   
 
A different initial state was
used by  
{\it Krasnopolsky et al.} 
(1999), who introduce and 
maintain throughout the simulation, a strong outflow on the 
outflow axis.
Otherwise, they choose an initial disk field distribution
that is the same as PP92, a mass flux density $\rho v_z \propto
r_o^{3/2}$, so that $k = const.$.  
The values of $B_r$ and $B_{\phi}$ at the
disk surface were not fixed in their simulations.
(This does not
guarantee that the disk or the boundary
at the base will remain  Keplerian over time).
The fixed  $B_{\phi}\propto r_0^{-1}$ profile adopted in the OP setup
ensures  exactly that - see the discussion around Eq~(3.46) in OPI).
To evade the  problems with the initial setup  mentioned above,
these authors continuously launch a cylindrical wind on the
axis. This imposed jet introduces  
currents which could significantly affect the stability and collimation 
of the disk wind.  Their disk mass loading would predict that the
disk wind should not be well collimated (see Section 2.4), whereas their simulation
does appear to collimate.  This suggests that their on-axis
jet may be playing a significant role in the simulation results.
   
\subsection{The role of mass loading}

The mass-load $k$, defined in equation~(8), 
can be established  
by  varying; the coronal density profile 
while keeping $\rho  v_{\rm p}$ constant  ({\it Anderson et al.}, 2005),  
the disk rotational profile 
since $v_{\rm p}= v_{\rm inj} v_{\phi}$ 
(e.g., {\it Vitorino et al.}, 2002),  
the distribution of the poloidal magnetic field 
on the disk (PRO), or the distribution of both the disk magnetic field
and the mass flow profile ({\it Fendt}, 2005).   

The prediction that the mass load determines the collimation of the jet
was tested in simulations by PRO.  
Fig.~1 shows that simulations with $\mu = 0, \ -0.25$ collimate
into cylinders, while the $\mu = -0.5$ case (PP92) transitions towards
the wide angle flow seen in the $\mu =-0.75$ simulation.  These
results confirm the theory laid out in Section 2.4. 
We also note that each of the simulations mentioned above results
in a unique rotation profile of the jet that might in principle
be observable.  Going from the potential case to the Pelletier-Pudritz
configuration, the radial profiles of the rotational velocity of the
jets scale as power laws 

\begin{equation}
v_{\phi}(r,\infty) \propto r^a
\end{equation} 
where 
$a = -0.76; \quad -0.66; \quad -0.46$ respectively.
The observation of a rotational profile would help
pick out a unique mass loading in this model (PRO).

The mass loading also affects the time-dependent behaviour
of jets.
{\it As the mass load is varied a transition from stationary, 
to periodic and sometimes to a discontinuous dying jet occurs}.  
For example, in some of the simulations   
it was found that low mass loads for jets lead to rapid, 
episodic behaviour while more heavily mass-loaded systems tend to 
achieve stationary outflow configurations  (see OPIII).

It is interesting that the 
simulations of {\it Anderson et al.}\ (2005) did not find any non-steady
 behavior for {\it low} mass loading (differently defined than
 equation (8)) but instead find an instability
 for {\it high} mass loading. They suggest that the origin
of the non-steady behavior for high mass loading is
 that the initially dominant toroidal field they impose could be subject
  to the kink instability (e.g., {\it Cao and Spruit}, 1994). The very large
  mass-loads, and large injection speeds ($v_{\rm inj}=(0.01,0.1)$)
   they use drive an instability probably related to excessive magnetic
   braking.  
Some of the main differences with OP are: 
(i) the use of the non-equilibrium set up of 
{\it Krasnopolsky et al.} which 
introduces a strong current along the axis that 
     could strongly affect the stability properties 
     of these outflows; and 
      (ii) the large injection
      speeds make the sonic surface too close to the disk 
      (the condition that $c_{\rm s} \ll v_{\rm K}$
       is not satisfied) and does
       not provide enough dynamical range for the gas launched
       from the disk to evolve smoothly.  Numerical instabilities 
        reminiscent of what was found by {\it Anderson et al.}\ (2005) were
	 observed  by OP and these 
        often disappear with high enough resolution
        (i.e. dynamical range). We suspect that this 
        type of instability will disappear 
	when a
        proper disk is included  in the simulations
        (e.g., \ {\it Casse and Keppens}, 2002).
 
\subsection{3-Dimensional simulations}

The stability of jets is one 
of the principal  remaining challenges in the theory.
It is well known that the purely toroidal field  
configurations that are used to help confine static, 
3-D, tokamak plasmas are  unstable (e.g., {\it Roberts}, 1967;
{\it Bateman}, 1980).  
The resulting kink, or helical 
($m=1$) mode instability derived from a 
3-D linear stability analysis is  powered by the 
free energy in the toroidal field, namely 
$B_{\rm \phi}^2 / 8  \pi$ ({\it Eichler}, 1993).  Why  are 
real 3-D jets so  stable over great distances in 
spite of the fact that they are probably threaded by strong toroidal fields?  

The 3D simulations needed to investigate the importance 
of kink modes are still rare.  
Early attempts include  
{\it Lucek and Bell} (1996)
and {\it Hardee and Rosen} (1999;
see also references therein) who performed 3-D simulations 
of ``equilibrium'' jets, and  found that these 
uniform, magnetized jet models remain 
Kelvin-Helmholtz (K-H) stable to low-order, 
surface helical and elliptical modes  ($ m = 1, 2$), provided that jets are on 
average sub-Alfv\'enic.  This is in accord with the prediction of linear 
stability analysis.  However, most configurations for jet simulations use 
rather {\it ad hoc\/} prescriptions for the 
initial toroidal field configuration so that it is 
difficult to assess how pertinent the results are to the case of a jet that 
establishes its own toroidal field as the jet is accelerated from the 
accretion disk.  In general, the available analytic and numerical results for 
the stability of simple jets show that the fastest growing modes are of K-H type.
These K-H instabilities are increasingly 
stabilized for super-Alfv\'enic jets, as $M_{\rm FM}$ is increased much beyond 
unity.  It is also generally known that sub-Alfv\'enic jets are stable.  Taken 
together, these results suggest that 3-D jets are the most prone to K-H 
instabilities a bit beyond their 
Alfv\'en surface, a region wherein their destabilizing 
super-Alfv\'enic character cannot yet be offset by the stabilizing effects 
engendered at large super FM numbers.  

\begin{figure}[t!]
\epsscale{1.0}
\begin{center}
\includegraphics[width=0.2\textwidth]{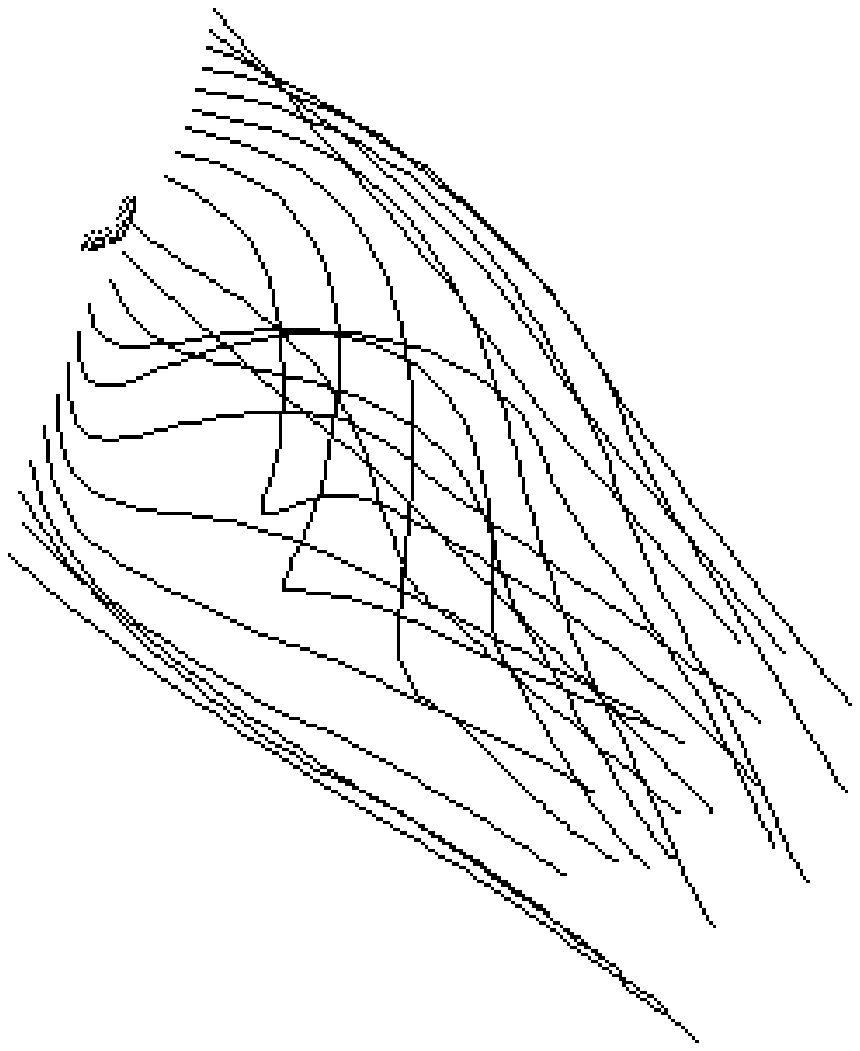}
\includegraphics[width=0.2\textwidth]{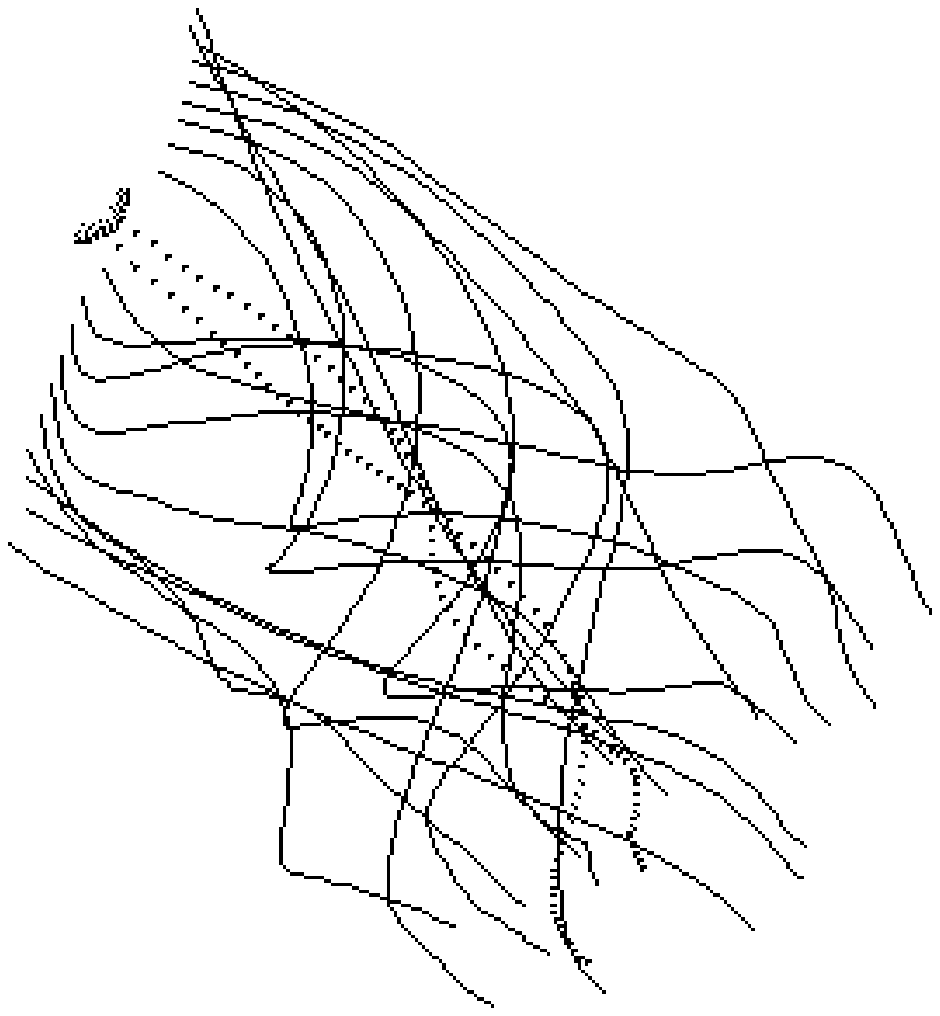}
\includegraphics[width=0.2\textwidth]{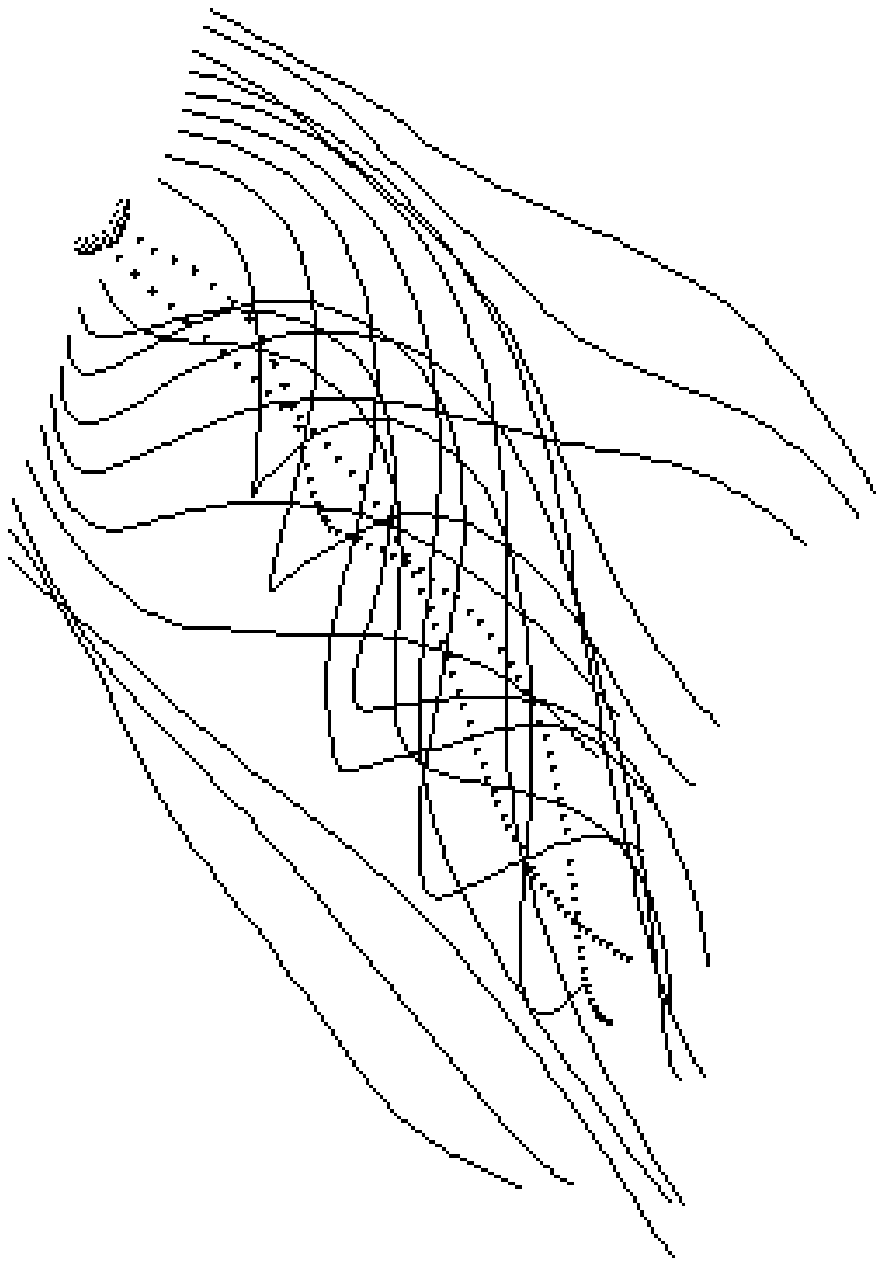}
\includegraphics[width=0.2\textwidth]{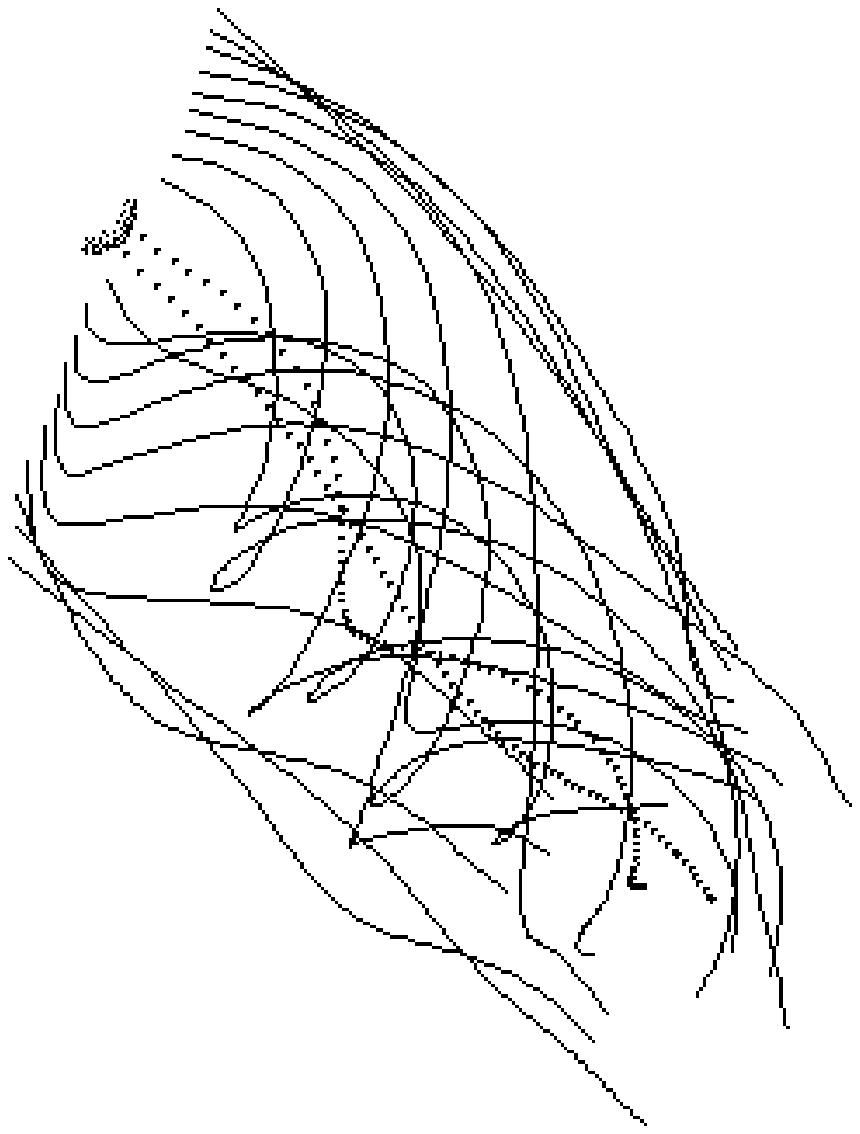}
\includegraphics[width=0.4\textwidth]{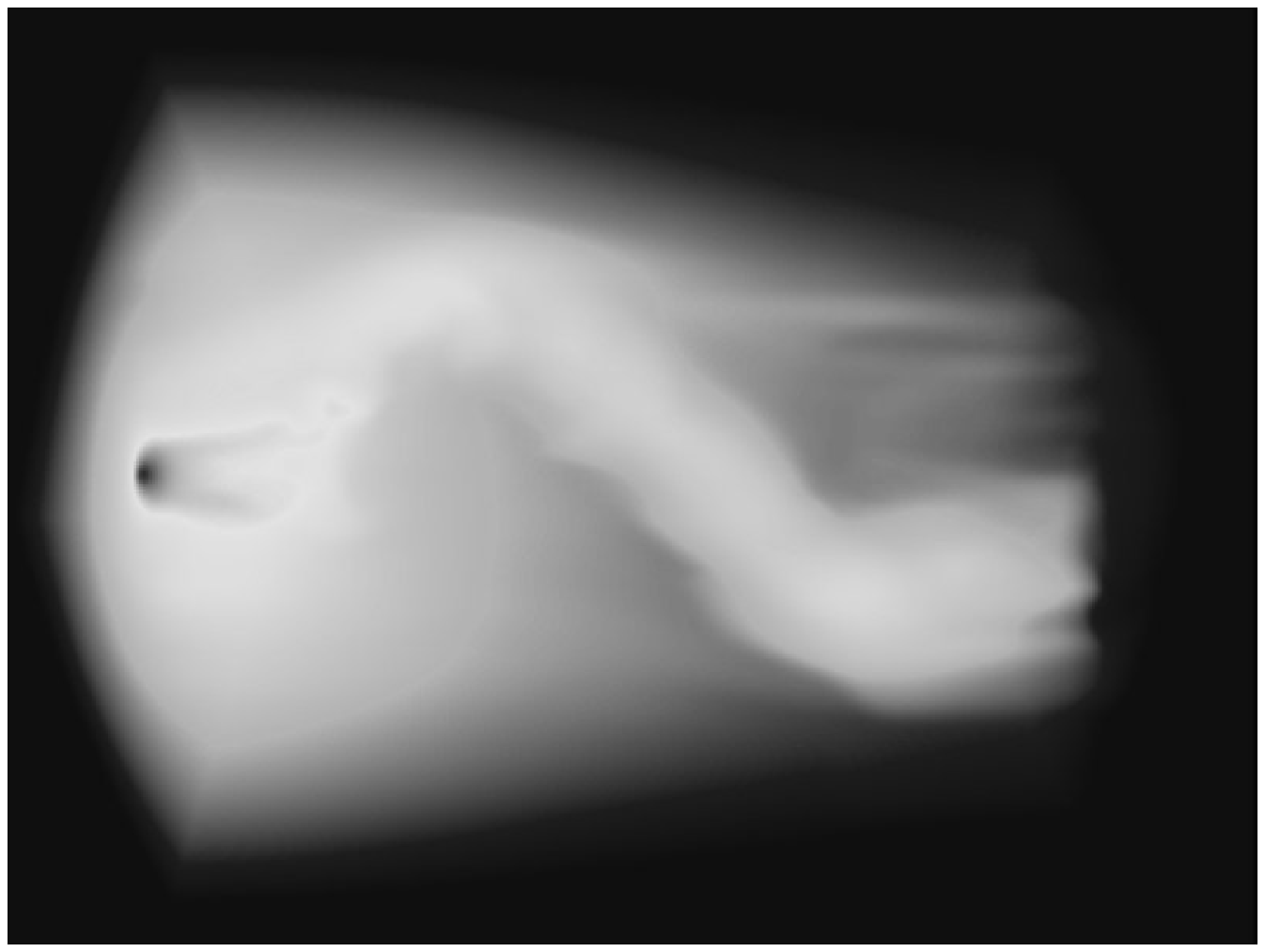}
\end{center}
\caption{\small Cylindrical Jets -- upper  panels show snapshots of 20 
magnetic field lines of the 3D simulations performed in OCP, 
 ** at $t=50, 130, 210$, and $240$.
The two central magnetic field lines (dotted lines) 
originate on the central compact object 
(illustrated by the semi-sphere to the left). 
The disk axis is along the diagonal of the frame 
(on a $45^\circ$ angle). Notice the violent jet behavior 
in the in-between frames when the kink mode appears. 
The jet eventually aligns itself with the original 
disk's axis acquiring a cylindrically collimated shape. 
  Corkscrew Jets -- The lower panel is a map of the jet column density at $t=320$
 for  a simulation with a  $v_{\phi}$ profile,
imposed at the accretion disk, that is different from the simulation 
  shown in the upper four panels (see OCP).
   The jet has settled into a quasi-steady state structure in the shape of a  
    ``corkscrew" (colors arranged from blue to red
  to represent low and high values of the density). 
The disk (not visible) 
is on the left-hand side of the image, and outflow is from left to right. 
[Adapted from OCP]. }
\label{fig:3d}
\end{figure}

There have, as yet, only been a few attempts 
to simulate 3D disk winds.
Contrary to what may be intuitive, 
it is inadvisable to perform these 3-D simulations in 
cylindrical coordinates. For one thing, 
special treatment must be given to the ``wedge zones" 
that abut the $z$-axis (no longer a symmetry axis in 3D), 
and velocities that pass through the $z$-axis pose a very 
difficult numerical problem. Secondly, even with such technical 
details in hand, plane waves are badly disrupted on passing 
through the $z$-axis, and this provides an undesirable bias 
to what should be an unbiased (no axis should be preferred over another) 
3D calculation. Fortunately, using Cartesian coordinates 
to simulate cylinders is feasible with careful 
setups as has been demonstrated in {\it Ouyed et al.}\ (2003; OCP); 
and {\it Ouyed} (2003).  These authors used a 3D version of the OPI setup.  

The central result of this study is 
that jets survive the threatening non-axisymmetric 
kink ($m=1$) mode.  The simulated jets maintain their long-term 
stability through a self-limiting  process wherein 
the average Alfv\'enic Mach number within the jet 
is maintained  to order unity.  This is accomplished in 
at least two ways: (i) poloidal  magnetic field 
is concentrated along the central axis of the jet forming a 
``backbone'' in which the Alfv\'en speed is sufficiently high to reduce the 
average jet Alfv\'enic Mach number to unity, and (ii)  
the onset of higher order 
Kelvin-Helmholtz ``flute'' modes ($m \ge 2$) reduce the efficiency with which 
the jet material is accelerated, and transfer kinetic energy of the outflow 
into the stretched, poloidal field lines of the distorted jet.  
This too has the
effect of increasing the Alfv\'en speed, and thus reducing the Alfv\'enic Mach 
number.  The jet is able to survive the onset of 
the more destructive $m=1$  mode in this way.  
It was further  discovered  that jets go into alternating 
periods of low and high activity as the disappearance of unstable modes in the 
sub-Alfv\'enic regime enables another cycle of 
acceleration to super-Alfv\'enic 
speeds. The period of the episodic (low vs. high) behavior uncovered
  in the case of the wobbling jet (upper four panels in Fig.~\ref{fig:3d})
   is of the order of $200~t_i$. For a typical young stellar
   object of mass $0.5M_{\odot}$ and for $r_{\rm i}\sim 0.05$ AU (i.e. $t_{\rm i}\sim 0.9$ day),
    this would correspond to a minimum period of roughly $180$ days. This
    is consistent with the temporal variations of the observed  YSOs outflow velocities which appears
     to occur on timescales of a few years ({\it Woitas et al.}, 2002).

Jets ultimately settle into relatively stable end-states involving either 
corkscrew (lower panel in Fig.~\ref{fig:3d}), or wobbling types of  
structure (upper four panels in Fig.~\ref{fig:3d}).
The difference 
in the two type of jets 
can be traced to the difference in the  $v_{\phi}$ profiles 
imposed at the accretion disk (see OCP for more details).
This trend has been uncovered in  3D simulations 
of large scale jets where  wiggled jets have also 
been observed  after the jets survived the 
destructive era ({\it Uchida et al.}, 1992; {\it Todo et al.}, 1993; 
{\it Nakamura et al.}, 2001; {\it Nakamura and Meier}, 2004).  
  For completeness, we should mention the 
recent similar simulations performed by {\it Kigure and Shibata} (2005)
which included
the disk self-consistently. These studies also show that non-axisymmetric
perturbations dominate the dynamics.
However, these simulations were only run for very short
times, 
never exceeding 2 inner-disk orbital periods due to 
numerical obstacles reported by the authors.  
In OCP simulations, it was shown that 
 the Cartesian grid (also used by {\it Kigure and Shibata}) induces
  artificial modes in the early stages of the simulations and only after
  many orbital periods that these become washed out before the real modes
   enter the dynamics.
Another suggested end-stated for 3D jets
is a minimum-energy
``Taylor state", wherein the jet maintains comparable poloidal and toroidal
 field components ({\it K\"onigl and Choudhuri}, 1985).  This alternative 
does not pertain to our simplified simulations since we do
not allow for  internal magnetic energy dissipation.  

In summary, 
a large body of different simulations 
have converged to show that jets; 
(i) are centrifugally disk winds; 
(ii) are collimated  by the ``hoop stress'' engendered by 
their toroidal fields; 
(iii) achieve two types of configuration depending on the 
mass loading that takes place in the underlying 
accretion disc -- those that achieve collimation 
towards a cylinder, and those that have a 
wide-angle structure; (iv) achieve two types of regimes depending on the 
mass loading that takes place in the underlying 
accretion disc -- those that achieve a stationary state, 
and those that are episodic;  
(v) achieve stability by a  combination of MHD mode 
coupling and ``back-bone" effect leading to a 
self-regulating mechanism which  saturates  the  
instabilities; and (vi) achieve different morphologies 
ranging from cylindrical wobbling to 
cork-screw structure, depending on the profile of 
$v_{\phi}$ imposed on the underlying accretion disk.

\section{\textbf{COUPLED DISK-JET EVOLUTION}}

While it is physically useful to regard the accretion disk 
as a boundary condition for the jet,
this ignores critical issues such as the self-consistent 
radial distribution of the mass loading, or of the 
threading magnetic field
across the disk.
These, and other degrees of freedom can be 
found by solving the
combined disk-outflow problem, 
to which we now turn.

Significant progress on the launch of disk winds has been
made by theoretical studies of a restricted 
class of self-similar 2D models 
(e.g.\, {\it Wardle and K\"onigl}, 1993; {\it Li}, 1995, 1996;
{\it Ferreira}, 1997; {\it Casse and Ferreira}, 2000).    
Since centrifugally driven outflows can occur for completely
cold winds (BP82), the models have examined both ``cold" and
``warm" (i.e., with some kind of corona) conditions.
In the former class, 
material
from some height above the disk midplane must move upwards and
eventually be accelerated outwards in a jet. Therefore, there must be a direct
link between the physics of magnetized disks, and the
origin of outflows.

Calculations show that in order to match the properties of jets
measured by {\it Bacciotti et al.}\ (2000), warm wind solutions
are preferred, wherein a disk corona plays a central role. 
In this situation, gas pressure imbalance will assist with
feeding the jet.  A warm disk corona is expected on general grounds because  
it is a consequence of the magneto-rotational instability (MRI), 
as the vertically resolved disk
simulations of {\it Miller and Stone} (2000) have shown.  

A semi-analytic model of the radial and vertical disk structure,
that includes self-consistently the outflow, has also been presented by
{\em Campbell} (2003).
These solutions demonstrate explicitly that the outflow contributes
to the loss of angular momentum in the disk by channeling it along
field lines into the outflow.
For self-consistent solutions to be possible, the turbulent Mach number
has to be between 0.01 and 0.1.

Models of outflows and jets generally assume ideal (non-resistive)
MHD.
However, inside the disk non-ideal effects must become important, because
the accreting matter would otherwise never be able to get onto the field
lines that thread the disk and connect it with the outflow. 
In recent two-dimensional models, {\em Casse and Keppens} (2002, 2004) assumed
a resistivity profile, analogously to the Shakura and Sunyaev prescription.
This assumes the presence of an underlying turbulence within the disk.
Only fairly large values of the corresponding $\alpha_{\rm SS}$ parameter
of around 0.1 have been used.
The subscript SS refers to {\em Shakura and Sunyaev} (1973), who were the
first to introduce this viscosity parameter.
In all cases the system evolves to a steady equilibrium.
Using resistive simulations in a different context,
{\it Kuwabara et al.} \ (2005) demonstrated that a substantial amount
of energy can be transported by Poynting flux if the poloidal field
falls off with distance no faster than $r^{-2}$.
Otherwise, the fast magnetosonic point is located closer to the
Alfv\'en point and the jet will be dominated by kinetic energy,
which is the case in the simulations of {\em Casse and Keppens} (2004).

The general stability of disk--outflow solutions is still being debated and the 
result may depend on the detailed assumptions about the model. 
In the solutions discussed here the accretion stress comes entirely from
the large scale magnetic field rather than some small scale turbulence.
As emphasized by {\em Ferreira and Casse} (2004),  
real disks have a turbulent viscosity just as
they have turbulent magnetic diffusivity or resistivity.
However, if the accretion stress does
come entirely from the large scale magnetic
field, the wind--driven accretion flows may be unstable
({\em Lubow et al.}, 1994).
{\em K\"onigl} (2004) has shown recently that 
there are in fact two distinct solution
branches--a stable and an unstable one.
He argues that real disk/wind systems
would correspond to the stable branch.

Finally, the idea of a gently flared accretion disk is likely to 
be merely the result of
the modeler's simplification rather than observational reality.
Indeed, accretion disks can be warped due to various instabilities which
can be driven by radiation from the central object (e.g., \ {\em Pringle}, 1996),
or, more likely, by the outflow itself ({\em Schandl and Meyer}, 1994).
If a system is observed nearly edge-on, a warp in the accretion disk
produces periodic modulations of the light curve.
As {\it Pinte and M\'enard} (2004) have demonstrated,
this may be the case in AA Tau.
Observations frequently reveal major asymmetries in bipolar outflows,
which may be traced back to the corresponding asymmetries in the
disk itself.
The possible causes for these asymmetries may be either an
externally imposed asymmetry such as one-sided heating by a
nearby OB association, or an internal symmetry breaking of the
disk-wind solution as a result increased rotation.
Examples of the latter are familiar from the study of mean field
dynamo solutions of accretion disks ({\it Torkelsson and Brandenburg}, 1994).

\subsection{Dead zones}

So far it has been assumed that the magnetic field is 
well coupled to the disk.  
This  is certainly valid for most of the envelope of an accretion disk,
which is
ionized by a combination of 
cosmic rays, as well as X-rays from the central YSO. 
However, the deeper layers of the disk are strongly shielded from 
these ionizing agents 
and the degree of ionization plumets. 

This dense layer, which encompasses the bulk of
the disk's column density, cannot maintain  
a sufficiently high electron 
fraction, and is referred to as the dead zone 
({\it Gammie}, 1996). It is the poorly ionized region   
within which the MRI 
fails to grow as a consequence of the diffusivity of the field.
Recent work of {\em Fleming and Stone} (2003) shows that, although the local
Maxwell stress drops to negligible values in the dead zones, the
Reynolds stress remains approximately independent of height and never
drops below approximately 10\% of the maximum Maxwell stress,
provided the column density in that zone is less that 10 times
the column of the active layers.
The non-dimensional ratio of stress to gas pressure is just
the viscosity parameter, $\alpha_{\rm SS}$.
{\it Fleming and Stone} (2003) find typical values of a few times $10^{-4}$ in
the dead zones and a few times $10^{-3}$ in the MRI-active layers. 
{\em Inutsuka and Sano} (2005) have questioned the very existence of
dead zones, and proposed that the turbulent dissipation in the disk
provides sufficient energy for the ionization.
On the other hand, their calculation assumes a magnetic Reynolds
number that is much smaller than expected from the simulations
(see {\it Matsumura and Pudritz}, 2006; MP06).

The radial extent of the dead zone depends primarily upon
the disk column density 
as well as effects of grains and radiation chemistry 
(e.g., {\it Sano et al.}, 2000; {\it Matsumura and Pudritz},
2005, MP05; and MP06).
Inside the dead zone the magnetic Reynolds number tends to be
below a certain critical value that is somewhere between 1 and 100
({\em Sano and Stone}, 2002), making MRI-driven turbulence impossible.
Estimates for the radial extent of the dead zone range from 
0.7--100 AU ({\it Fromang et al.}, 2002), 
to 2--20 AU in calculations by {\it Semenov et al.}\ (2004).
For {\it Chiang et al.}\ (2001) models of disks that are well 
constrained by the observations, and whose surface density 
declines as $\Sigma \propto r_0^{-3/2}$, MP05 find that the 
extent of the dead zone is robust -- extending out
to 15 AU typically, and is fairly independent 
of the ionizing environment of the disk.  

For smaller radii, thermal and UV ionization are mainly responsible for
sustaining some degree of ionization.
The significance of the reduced value of $\alpha_{\rm SS}$ in the dead zones
is that it provides a mechanism for stopping the inward migration of
Jupiter-sized planets (MP05, MP06).
Jets can still be launched from the well-coupled surface layer
above the dead zone
(e.g., {\em Li}, 1996;
{\em Campbell}, 2000).

When the MRI is inactive in the body of the disk, 
alternative mechanisms of angular momentum
transport are still possible. 
In protostellar disks there are probably at least two other mechanisms
that might contribute to the accretion torque:
density waves ({\em R\'o\.zyczka and Spruit}, 1993),
and the interaction with other planets in the disk
({\em Goodman and Rafikov}, 2001). 

A more controversial alternative is to drive turbulence by a nonlinear
finite amplitude instability ({\it Chagelishvili et al.}, 2003).
While {\it Hawley et al.}\ (1999) have given general arguments against this
possibility, {\it Afshordi et al.}\ (2005) and other groups have continued
investigating the so-called bypass to turbulence.
The basic idea is that successive strong transients can maintain a
turbulent state in a continuously excited manner.
{\em Lesur and Longaretti}\ (2005) have recently been able to quantify more
precisely the critical Reynolds number required for instability.
They have also highlighted the importance of pressure fluctuations that
demonstrate that the general argument by {\em Balbus et al.}\ (1996)
is insufficient.

Finally, the role of vertical (convectively stable) density stratification
in disks and the possibility of the so-called strato-rotational instability
has been proposed as a possible mechanism for disk turbulence
({\it Dubrulle et al.}, 2005).
This instability was recently discovered by {\it Molemaker et al.}\ (2001) and
the linear stability regime was analyzed by {\it Shalybkov and R\"udiger} (2005).
However, the presence of no-slip radial boundary conditions that are
relevant to experiments and used in simulations are vital.
Indeed, the instability vanishes for an unbounded regime, making it
irrelevant for accretion disks ( {\it Umurhan}, 2006).

\subsection{Advected vs.\ dynamo generated magnetic fields}

\newcommand{\meanBB}{\overline{\mbox{\boldmath $B$}}}
\newcommand{\AU}{\,{\rm AU}}
\newcommand{\K}{\,{\rm K}}

One of the central unresolved issues in disk wind theory is 
the origin of the threading magnetic field.  Is it
dragged in by the gravitational collapse of an original
magnetized core, or is it generated {\it in situ} by a disk-dynamo
of some sort?

The recent direct detection of a rather strong, true disk field
of strength 1 kG at 0.05 AU in FU Ori, provides new and strong support
for the disk wind mechanism ({\it Donati et al.} 2005).
The observation technique uses 
high-efficiency
high-resolution spectropolarimetry and holds
much promise for further measurements. 
This work provides excellent evidence that distinct fields
exist in disks in spite of processes such as ambipolar
diffusion which might be expected to reduce 
them.

{\em 4.2.1 Turbulence effects.}
Accretion disks are turbulent 
in the well-coupled lower corona and beyond.  They are also intrinsically
three-dimensional and unsteady.
Questions regarding the stability of disks concern therefore only the mean
(azimuthally averaged) state, where all turbulent eddies are averaged out.
The averaged equations used to obtain such solutions incorporate a
turbulent viscosity.
In this connection we must explain that the transition from a
macroscopic viscosity to a turbulent one is more than just a change
in coefficients, because one is really entering a new level of
description that is uncertain in many respects.
For example, in the related problem of magnetic diffusion it has been
suggested that in MRI-driven turbulence, the functional form of
of turbulent magnetic diffusion may be different, such that it operates
mainly on the system scale and less efficiently on smaller scales
({\em Brandenburg and Sokoloff}, 2002).
It is therefore important to use direct simulations to investigate the
stability in systems that might be unstable according to a mean field
description.
There is no good example relevant to protostellar disks, but related
experience with radiatively dominated accretion disks, where
it was found that the
so-called Lightman--Eardly instability might not lead to the breakup of the
disk ({\em Turner et al.}, 2004), should provide enough reason to treat the
mean-field stability problem of protostellar disks with caution.
Also, one cannot exclude that circulation patterns found in simulations
of the two-dimensional mean-field equations (e.g., {\em Kley and Lin}, 1992)
may take a different form
if the fully turbulent three-dimensional problem was solved.

The turbulent regions of  disks will generally be
capable of dynamo action -- wherein an ordered field is generated
by feeding on the energy that sustains the turbulence.
These two aspects are in principle interlinked,
as has been demonstrated some time
ago using local shearing box simulations.
These simulations show not only that the magnetic field is generated from
the turbulence by dynamo action, but also that the turbulence itself is
a consequence of the magnetic field via the MRI
({\em Brandenburg et al.}, 1995; {\em Stone et al.}, 1996).

In comparing numerical dynamos with observational reality, it is
important to distinguish between two different types of dynamos:
small scale and large scale dynamos
(referring primarily to the typical physical scale of the field).
Both types of dynamos have in general a turbulent component,
but large scale dynamos have an additional component
on a scale that is larger than the typical scale of the turbulence.
Physically, this can be caused by the effects of anisotropies,
helicity, and/or shear.
These large scale dynamos are amenable to mean field modeling (see below).
Small scale dynamos, on the other hand, tend to be quite prominent
in simulations, but they are not described by mean field models.
However, small scale dynamos could be an artefact of using unrealistically
large magnetic Prandtl numbers in the simulations.
Protostellar disks have magnetic Prandtl numbers, $\nu/\eta$,
around $10^{-8}$, so the viscosity of the gas, $\nu$, is much smaller
than the diffusivity of the field, $\eta$;
see Table~1 of {\em Brandenburg and Subramanian} (2005).
High magnetic Prandtl numbers imply that the field
is advected with the flow without slipping too much, whereas for low numbers
the field readily slips significantly with respect to the flow.
At the resistive scale, where the small scale dynamo would
operate fastest, the velocity is still in its inertial range where
the spatial variation of the velocity is much rougher than for unit
magnetic Prandtl number ({\em Boldyrev and Cattaneo}, 2004).
This tends to inhibit small scale dynamo action
({\it Schekochihin et al.}, 2005).
In many simulations, especially when a subgrid scale prescription is
used, the magnetic Prandtl number is effectively close to unity.
As a consequence, the production of small scale field may be exaggerated.
It is therefore possible that in real disks large scale dynamo
action is much more prominent than what is currently seen in simulations.

In mean field models, only the large scale field is modeled.
In addition to a turbulent magnetic diffusivity that emerges
analogously to the turbulent viscosity mentioned above,
there are also non-diffusive contributions such as the
so-called $\alpha$ effect
wherein the mean electromotive force can 
acquire a component
parallel to the mean field, $\alpha_{\rm dyn}\meanBB$;
see {\em Brandenburg and Subramanian} (2005) for a recent review.
For symmetry reasons, $\alpha_{\rm dyn}$ has on average opposite signs
above and below the midplane.
Simulations of {\em Ziegler and R\"udiger} (2000) confirm the
unconventional negative sign of $\alpha_{\rm dyn}$ in the upper disk
plane, which was originally found in {\em Brandenburg et al.}\ (1995).
This has implications for the expected field parity which
is expected to be dipolar (antisymmetric about the midplane); see
{\it Campbell et al.}\ (1998) and {\it Bardou et al.}\ (2001).
Although a tendency toward dipolar fields is now seen in some
global accretion disk simulations ({\em De Villiers et al.}, 2005),
this issue cannot as yet be regarded as conclusive.

{\em 4.2.2 Outflows from dynamo-active disks.}
Given that the disk is capable of dynamo action, what are the
relative roles played by ambient and dynamo-generated fields?
This question has so far only been studied in limiting cases.
The first global simulations were axisymmetric, in which case dynamo
action is impossible by the {\em Cowling} (1934) theorem.
Nevertheless, such simulations have demonstrated quite convincingly
that a jet can be launched by winding up the ambient field and driving
thereby a torsional Alfv\'en wave ({\em Matsumoto et al.}, 1996).
These simulations are scale invariant and, although they were originally
discussed in the context of active galaxies, after appropriate
rescaling, the same physics applies also to stellar disk outflows.

When the first three-dimensional simulations became available an
immediate issue was the demonstration that MRI and dynamo action
are still possible ({\em Hawley}, 2000).
Although the simulations were applied to model black hole accretion
disks, many aspects of such models are sufficiently generic
that they carry over to protostellar disks as well.
These simulations showed that the dynamo works efficiently
and produces normalized accretion stresses ($\alpha_{\rm SS} \approx 0.1$)
that exceed those from local simulations ($\alpha_{\rm SS} \approx 0.01$).
The global simulations did not at first seem to show any signs of
outflows, but this was mainly a matter of looking at suitable
diagnostics, which emphasizes the tenuous outflows rather than the
much denser disk dynamics ({\em De Villiers et al.}, 2005).
Although these simulations have no ambient field, a large scale field
develops in the outflow region away from the midplane.
However, it is difficult to run such global simulations for long enough to
exclude a dependence of the large scale field on the initial conditions.
The outflows are found to be uncollimated such that most of the
mass flux occurs in a conical shell with half-opening angle of
$25$--$30^\circ$ ({\em De Villiers et al.}, 2005).

These very same properties are also shared by mean field simulations,
where the three-dimensional turbulent dynamics is
modeled using axisymmetric models ({\em von Rekowski et al.}, 2003).
In these simulations (Fig.~3), where the field is
entirely dynamo-generated, there is an uncollimated outflow with most of
the mass flux occurring in a conical shell with a half-opening angle of about
$25$--$30^\circ$, just like in the three-dimensional
black hole simulations.
The field strength generated by dynamoc action in the disk is found to scale as 
$B_p \propto r^{-2}$.  This is very similar to the scaling 
for poloidal field strength found in collapse simulations
(see {\it Banerjee and Pudritz}, 2006). 

\begin{figure}[t!]
\begin{center}
\includegraphics[width=.99\columnwidth]{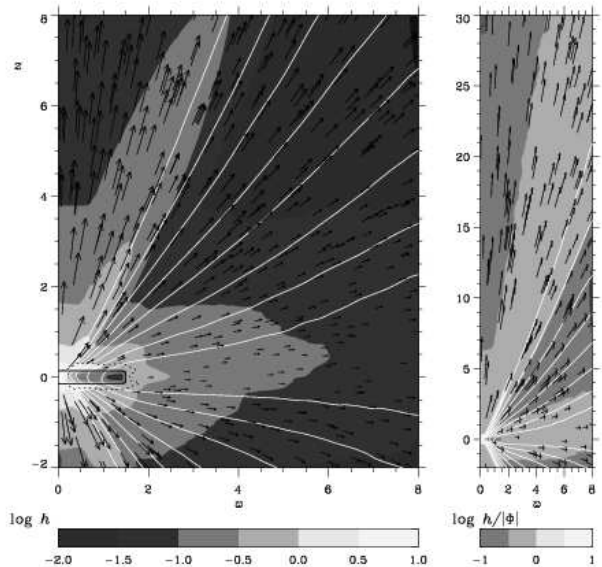}
\end{center}\caption[]{
Outflow from a dynamo active accretion disk driven by a combination
of pressure driving and magneto-centrifugal acceleration.
The extent of the domain is $[0,8]\times[-2,30]$ in nondimensional
units (corresponding to about $[0,0.8]\AU\times[-0.2,3]\AU$ in
dimensional units).
Left panel: velocity vectors, poloidal magnetic field lines
and gray scale representation of the specific enthalpy $h$
in the inner part of the domain.
Right panel: velocity vectors, poloidal magnetic field lines and
normalized specific enthalpy $h/|\Phi|$ in the full domain.
[Adapted from {\em von Rekowski et al.}\ (2003)].
}\label{FRRRRun3c}\end{figure}

The conical outflows discussed above
have tentatively been compared with the observed conical
outflows inferred for the BN/KL region in the Orion nebula out to a
distance of 25--60 AU from its origin ({\em Greenhill et al.}, 1998).
One may wonder whether collimated outflows are only possible
when there is an ambient field (see Section 3).
This conclusion might be consistent with observations of the Taurus-Auriga
molecular cloud.
{\em M\'enard and Duch\^ene} (2004) found that,
although T Tauri stars are oriented randomly with respect to the
ambient field, there are no bright and extended outflows when the
axis is perpendicular to the ambient field.
Note that {\em Spruit and Uzdensky} (2005) have argued that the efficiency
of dragging an ambient field toward the star have been underestimated in
the past, provided the field segregates into many isolated flux bundles.

In summing up this section
we note that all magnetic disk models produce outflows
of some form.
However, the degree of collimation may depend on the possibility of
an ambient field. 
An important ingredient that seems to have been ignored in all combined
disk-jet models is the effect of radiation.
There are also many similarities between
the outflows in models with and without explicitly including the disk 
structure.
Although these simulations are non-ideal (finite viscosity and
resistivity), the various Lagrangian invariants (mass loading parameter,
angular velocity of magnetic field lines, as well as the angular momentum
and Bernoulli constants) are still very nearly constant along field lines
outside the disk.

\section{\textbf{JETS AND STAR-DISK INTERACTION}}

In this section we review the recent development concerning the interaction
of young stellar magnetospheres with the surrounding accretion disk and the
contribution that these processes may have to the formation of jets.

Perhaps the most important clue that star-disk interaction is
an important physical process 
comes from the observation that 
many YSOs 
are observed to spin at only 
a small fraction ($ \simeq 10\% $) of their
break-up speed (e.g., {\it Herbst et al.}, 2002).  
They must therefore 
undergo a significant spin-down torque 
that almost cancels the strong spin-up that 
arising from gas accretion from the inner edge of the Keplerian
disk.

Three possible spin-down torques have been
proposed:
(i) disk-locking -- a magnetospheric connection between
the disk and the star which dumps accreted 
angular momentum back out into the disk beyond the 
co-rotation radius 
({\it K\"onigl} 1991); (ii) an  
X-wind --  a 
centrifugally driven outflow launched from the very inner 
edge of the accretion disk which intercepts 
the angular momentum destined for the star
(see chapter by {\it Shang et al.}); and (iii) 
an accretion-powered stellar wind -- 
in which the slowly rotating, central star  
drives a massive ($\dot M_{\rm w,*} \simeq 0.1 \dot M_{\rm a}$), stellar wind 
that carries off the bulk of the angular momentum as
well as a fraction of the energy, that is deposited
into the photosphere by magnetospheric accretion 
({\it Matt and Pudritz}, 2005b).
The last possibility implies 
that the central object in an accretion 
disk also drive an outflow, which is in addition to the disk wind. 
Separating the disk wind from the outflow from the central
object would be difficult, but the shear layer that 
separates them might generate instabilities and shocks
that might be diagnostics.  

The strength of the star-disk coupling 
depends on the strength of the
stellar magnetic field.
Precise stellar magnetic field measurements exist for nine 
TTSs, 
while for seven other stars statistically significant fields have been found 
(see {\em Symington et al.}, 2005b and references therein).
Zeeman circular-polarization measurements 
in He~I of several classical TTs give 
direct evidence of kG magnetic fields with 
strong indication of considerable field
variation on the time scales of years ({\em Symington et al.}, 2005b).
For a sample of stars a magnetic field strength up to 4\,kG fields could be 
derived (DF Tau, BP Tau), as for others, notably the jet source DG Tau,
no significant field could be detected at the time of observation.

\subsection{Basic processes}

{\em 5.1.1 Variable accretion from a tipped dipole.}
A central dipolar field inclined to the rotation axis of star and disk
may strongly disturbs the axisymmetry of the system.
Photometric and spectroscopic variability studies of AA\,Tau 
give evidence for time-dependent magnetospheric accretion on time
scales of the order a month.
Monte-Carlo modeling shows that the observed photo-polarimetric 
variability may arise by warping of the disk that is induced
by a tipped  magnetic dipole
({\em O'Sullivan et al.}, 2005).
More general investigations of the warping process by numerical simulations
show that the warp could evolve into a steady state precessing rigidly
({\em Pfeiffer and Lai}, 2004). 
Disks can be warped by the magnetic torque that arises 
from the a slight misalignment 
between the disk and star's rotational axis ({\em Lai} 1999).
Disk warping may also operate in the absence of a stellar
magnetosphere since it can be induced 
by the interaction between a large-scale magnetic
field that is anchored in the disk, 
and the disk electric current. This leads to a warping
instability and to the retrograde precession 
of magnetic jets/outflows ({\em Lai}, 2003).

Three-dimensional radiative transfer models
of the magnetospheric emission line profile 
based on the warped disk density and velocity 
distribution obtained by numerical MHD 
simulations give gross agreement with 
observations with a variability somewhat larger 
than observed ({\em Symington et al.}, 2005a).

\begin{figure}[t!]
\epsscale{1.0}
\plotone{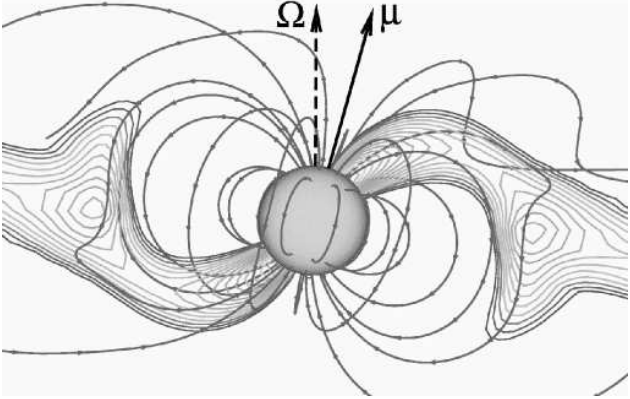}
\caption{\small Slice through a funnel stream. 
		  Density contours from 
                 $\rho = 0.2$ to $\rho = 2.0$.
                  The density of the disk corona is $\rho = 0.01-0.02$
                  Selected magnetic field lines.
                  The rotational axis and the magnetic moment are indicated
                  as ${\bf\Omega}$ and ${\bf \mu}$.
                 [Adapted from {\em Romanova et al.} (2004).]
          }  
\end{figure}

{\em 5.1.2 Magnetic flux.}
Compared to the situation of a pure disk magnetic field, the magnetic field of
the star may add substantial magnetic flux to the system.
For a polar field strength $B_0$ and a stellar radius $R_{\star}$,
the large scale stellar dipolar field (ignoring angular variations)
\begin{equation}
B_{\rm p,\star} (r) \simeq 40\,{\rm G} \left(\frac{B_0}{1\,{\rm kG}}\right)
                                       \left(\frac{r}{3\,R_{\star}}\right)^{-3}
\end{equation}
has to be compared to the accretion disk poloidal magnetic field 
which is provided
either by a disk dynamo or by advection of ambient interstellar field,
\begin{eqnarray}
\label{eq_beq}
B_{\rm p,disk} & < & B_{\rm eq}(r) = 20\,{\rm G}\,\alpha^{-1/2}
\left(\frac{\dot{M}_{\rm a}}{10^{-6}\,M_{\odot}\,{\rm yr}^{-1}} \right)^{1/2}
\nonumber
\\ 
 \phantom{B} & \, & 
\cdot
\left(\frac{M_{\star}}{M_{\odot}}\right)^{1/4}
\left(\frac{H/r}{0.1}\right)^{-1/2}
\left(\frac{r}{10\,R_{\odot}}\right)^{-5/4}.
\end{eqnarray}
where $B_{\rm eq}$ is the equipartition field strength in the disk.
This flux will not remain closed, but will inflate and open up as
magnetic field lines are sheared and extract gravitational
potential energy from the accreting flow 
(e.g., {\it Uzdensky et al.}, 2002; {\it Matt and Pudritz}, 2005a).
These field lines therefore effectively become a disk field,
and therefore follow the processes of disk wind production
we have already discussed.

The additional Poynting flux that threads
the disk may assist the jet launching by MHD forces and
may serve as an additional energy reservoir for the kinetic
energy of the jet implying greater asymptotic jet speed 
(Michel scaling; {\em Michel}, 1969; {\em Fendt and Camenzind}, 1996).

{\em 5.1.3 Disk locking vs.\ stellar winds.}
The spin of the star will depend on how angular momentum
arriving from the disk, is dealt with.
In the magnetic ``disk locking" picture, 
the threading field of the disk 
will re-arrange the global angular momentum budget.
The torque on the star by the accretion of disk matter is 
\begin{equation}
\tau_{\rm a} = \dot{M}_{\rm a} \left(G M_{\star} r_{\rm in}\right)^{1/2}
\end{equation}
(e.g., {\it Matt and Pudritz}, 2005a),
with the disk accretion rate $\dot{M}_{\rm acc}$, the stellar mass $M_{\star}$
and the disk inner radius $r_{\rm in}$ inside the corotation radius 
$r_{\rm co} = (G M_{\star} )^{1/3} \Omega_{\star}^{-2/3}$.
On the other hand if ``disk locking" is present, the stellar rotation may be
decelerated by the magnetic torque due to stellar field lines connecting the
star with the accretion disk outside $R_{\rm co}$.
The differential magnetic torque acting on a disk annulus of $dr$ width is
\begin{equation}
d\tau_{\rm mag} =  r^2 B_{\phi} B_{\rm z} dr.
\end{equation}
While $B_{\rm z}$ in principle follows from assuming a central dipolar field,
the induction of toroidal magnetic fields is model dependent 
(disk resistivity, poloidal field structure).
This is why recent numerical simulations of
dipole-disk interaction that simultaneously 
evaluate the poloidal and toroidal field 
components have become extremely valuable.

If indeed the star loses angular momentum to the disk (this is not yet
decided by the simulations, see below), both disk accretion and jet formation
are affected.
In order to continue accretion, excess angular momentum has to be removed from
the accreting matter. 
A disk jet can be an efficient way to do this, as has been
supposed in the X-wind picture. 

The central stellar magnetic field may launch a strong stellar wind
to rid itself of the accreted angular momentum.
Such an outflow will interact with 
the surrounding disk wind. 
If true, observed YSO jets may consist of two components -- the
stellar wind and the disk wind, with strength depending on intrinsic
(yet unknown) parameters.
The stellar wind (open field lines of stellar magnetosphere) 
exerts a spin-down torque upon the star of magnitude
\begin{eqnarray}
\dot{M}_{\rm wind, \star}\Omega_{\star} r_{\rm A}^2  = 
3\times 10^{36} \,\frac{\rm g\,cm^2}{\rm s\,yr}     
\left(\frac{\Omega_{\star}}{10^{-5}\,{\rm s^{-1}}}\right) \\ \nonumber
\cdot \left(\frac{r_{\rm A}}{30 R_{\odot}}\right)^2
\left(\frac{\dot{M}_{\rm wind, \star}}{10^{-9} M_{\odot} {\rm yr^{-1}}}\right)
\end{eqnarray}

As a historical remark we note that the model topology of dipole-plus-disk 
field were introduced for protostellar jet formation more then 20
years ago in the MHD simulations of {\it Uchida and Shibata} (1984).
Further investigations considering the detailed physical processes 
involved in disk truncation and channeling the matter along the dipolar 
field lines by
{\em Camenzind} (1990), {\em K\"onigl} (1991) and {\em Shu et al.}\ (1994)
resulted in a breakthrough of these
ideas to the protostellar jet community.

\subsection{Numerical simulations of star-disk interaction}

The numerical simulation of the magnetospheric star-disk interaction is 
technically most demanding since one must treat
a complex model geometry in combination with 
strong gradients in magnetic field strength, density and resistivity.
In general, this may imply a large variation in physical times scales for
the three components of disk, jet, and magnetosphere, which all have to be 
resolved numerically.
Essentially, numerical modeling of the star-disk interaction  
requires a {\em diffusive and viscous MHD code including radiative transfer}.
In addition, realistic models for reconnection processes and disk opacity
are needed.
Then, such a code has to run with {\em high spatial and temporal resolution}
on a global scale. 

Compared to the situation about a decade ago when there was still only 
{\em Uchida and Shibata}'s (1984) initial (though ten years old) simulation 
available, 
today huge progress has been made with several groups (and also codes) competing 
in the field.
Early simulations were able to follow 
the evolution only for a few rotations of
the inner disk (note that 100 rotations at 
the co-rotation radius
correspond to 0.3 Keplerian rotations at 10 co-rotation radii) 
({\em Hayashi et al.}, 1996; {\em Miller and Stone}, 1997; 
{\em Goodson et al.}, 1997). 
Among the problems involved is the initial condition of the simulation,
in particular the nature of the disk model which could be numerically treated.
A steady initial corona will strongly shear with the Keplerian disk leading to
current sheet (thus pressure gradients) along the disk surface. 
If the simulations run long enough, this could be an intermittent feature.
However, the danger exists that the artificial current sheet will fatally
destroy the result of the simulation.
Applying a Shakura-Sunyaev disk for the initial disk structure,
the code should also consider $\alpha_{SS} $ viscosity. 
Otherwise the initial disk evolution is not self-consistent 
(see, e.g., {\em Goodson et al.}, 1997).

The next step is to increase the grid resolution and to redo the
axisymmetric simulations.
{\em Goodson et al.}\ (1999) and {\em Goodson and Winglee} (1999)
were able to treat 
several hundreds of (inner) disk rotations on a global grid of 50 AU extension.
The main result of these simulations is a two component flow consisting of 
a fast and narrow axial jet and a slow disk wind,
both launched in the inner part of the disk ($r < 1\,{\rm AU}$).
An interesting feature is that the narrow axial jet is actually a collimation
in density and not in velocity.
Close to the inner disk radius repetitive reconnection processes are seen 
on time scales of a couple of rotation periods.
The dipolar field inflates and an expanding current sheet builds up.
After field reconnection along the current sheet, the process starts all over 
again.
The oscillatory behaviour leads to the ejection of axial knots.
On average the central dipolar magnetosphere remains present and loaded with 
plasma.
Forbidden emission line intensity maps of these simulations have been 
calculated and allow for direct comparison with the observations.
However, the numerically derived time and length scales of axial knots were 
still too different from what is observed.
The magnetospheric origin of jets (stellar dynamo) in favor of a pure disk
origin (``primordial field") has also been stressed by
{\em Matt et al.}\ (2002).

In order to be able to perform long-term simulations of dipolar magnetospheres
interacting with the accretion disk, 
{\em Fendt and Elstner} (1999, 2000) 
neglected the evolution of the disk structure 
and instead assume that the disk is a fixed boundary condition for the outflow
(as in OPI). 
After 2000 rotations a quasi-steady state was 
obtained with a two-component outflow 
from disk and star.
The outflow expands almost radially without signature of collimation on the 
spatial scale investigated ($20\times20$ inner disk radii).
One consequence of this very long simulation is that the axial 
narrow jet observed in other simulations 
might be an intermittent feature 
launched in the early phase of the simulation
as the initial field is reconfigured to a new equilibrium. 
The axial outflow in this simulations is massive but slow, but tends to 
develop axial instabilities for a lower mass loading.
Clearly, since the disk structure 
was not been taken into account,
nothing could be said about the launching process 
of the outflow out of the disk.

In a series of ideal MHD simulations Romanova and 
collaborators succeeded in working out
a detailed and sufficiently stable numerical model 
of magnetospheric disk interaction.
They were the first to simulate the axisymmetric 
funnel flow from the disk inner radius
onto the stellar surface ({\em Romanova et al.}, 2002) 
on a global scale ($R_{\max}= 50 R_{\rm in}$) and for a 
sufficiently long period of time
(300 rotations) in order to reach a steady state in the accretion funnel.
The funnel flow with free-falling material 
builds up as a gap opens up between
disk and star. 
The authors further investigated the angular
momentum exchange due to the disk
``locking" by the funnel flow.
Slowly rotating stars seem to break the inner disk region and spin up, while
rapid rotators would accelerate the inner disk region to
super-Keplerian velocity and slow down themselves.
Only for certain stellar rotational periods a ``torque-less" accretion could
be observed.
Strong outflows have not been observed for the parameter space investigated,
probably due to the matter dominated corona 
which does not allow for opening-up
the dipolar field.

Further progress has been achieved 
extending these simulations to three dimensions
({\em Romanova et al.}, 2003, 2004). 
For the first time it has been possible to 
investigate the interaction of an inclined
stellar dipolar magnetosphere with the surrounding disk.
For zero inclination axisymmetric funnels 
are found as in the axisymmetric simulations.
For non-zero inclination the accretion splits in 
two main streams following the closest
path from the disk to the stellar surface
as is seen in Fig.~4. 
Magnetic braking changes the disk structure 
in several ways -- its density structure
(gaps, rings), its velocity structure 
(sub-Keplerian region) and its geometry (warping).
The slowly rotating star is spun up by accreting 
matter but this acceleration does
only weakly depend on inclination.
For completeness we should note that for 
these simulations a clever approach for the
numerical grid has been applied, the 
``cubed grid" which does not obey the singular
axes as for spherical coordinates. 

The star-disk coupling by the stellar magnetosphere was also investigated 
by {\em K\"uker et al.}\ (2003). These simulations have been performed
in axisymmetry, but an advanced disk model has been applied.
Taking into account $\alpha$-viscosity, 
a corresponding eddy magnetic diffusivity
and radiative energy transport, the code enables the authors to treat 
a realistic {\em accretion} disk in their MHD simulations. 
As a result the simulations could be advanced to very long physical time steps
into a quasi stationary state of the disk evolution.
The authors show that the commonly assumed 
1000\,G magnetosphere is not sufficient
to open up a gap in the disk.
Unsteady outflows may be launched outside of 
the corotation radius with mass loss
rates of about 10 \% of the accretion rate.
The authors note that they were unable 
to detect the narrow axial jet seen in other
publications before (in agreement with {\em Fendt and Elstner}, 2000).
Magnetic braking of the star may happen in some cases, but is overwhelmed by
the accretion torque still spinning up the star (however, these simulations 
treat the stellar surface  as inner boundary condition).

Using a different approach, 
the boundary condition 
of a stellar {\em dipolar} magnetosphere 
or a large scale disk field were relaxed. 
This allows for the 
interaction of a stellar dipole with a
{\em dynamo}-generated disk magnetic field 
({\em von Rekowski and Brandenburg}, 2004),
or the evolution of a dynamo-generated stellar 
field affected by a surrounding disk
with its own disk dynamo ({\em von Rekowski and Brandenburg}, 2006)
to be studied.
The results of this work are in agreement with previous studies, 
and directly prove that a disk
surrounding a stellar magnetosphere may actually 
develop its own magnetic field
strong enough to launch MHD outflows. 
In agreement with results by {\em Goodson and Winglee} (1999),
accretion tends to be unsteady and alternating between connected
and disconnected states.
Even for realistic stellar fields of several hundred gauss the magnetic
spin-down torque is insufficient to overcome the spin-up torque
from the accretion flow itself.
As shown by the authors, angular momentum exchange is complex and
may vary in sign along the stellar surface, braking some parts of the star and
accelerating others.
The dynamo-generated stellar field may reach up to 750\,G and switches in time between
dipolar and quadrupolar symmetry, 
while the dynamo-generated 50\,G disk field is of dipolar symmetry.
In general, the dynamo-generated stellar field is better suited to drive a stellar
wind. 
The simulations show that for these cases stellar wind braking is more 
efficient than braking by the star-disk coupling, confirming the results
of {\it Matt and Pudritz} (2005a).

Recent studies of torque-less accretion ({\em Long et al.}, 2005) compare 
cases of a (i) weak stellar magnetic field within a dense corona and (ii) 
a strong field in a lower density corona.
They investigate the role of quasi-periodic field line reconnection 
in coupling the disk and stellar fields. 
Unlike previous works, these authors conclude that
magnetic interaction effectively locks the stellar rotation to about 10\% the 
break-up velocity, a value which actually depends on the disk
accretion rate and the stellar magnetic moment.
While they correctly stress the 
importance of dealing with the exact balance between 
open and closed field lines and their corresponding angular momentum flux,
the exact magnetospheric structure in the innermost region will certainly
depend on the resistivity (diffusivity) of the disk and corona material.
This is, however, not included in their 
treatment as an ideal MHD code has been applied.

We summarize this section noting that  
tremendous progress has occurred in the numerical
simulation of the star-disk interaction.
The role of  numerical simulations is pivotal in this area because the mechanisms involved 
are complex, strongly interrelated, and often highly time-dependent.
It is fair to say that numerical simulations of the star-disk interaction have not yet 
shown the launching of a jet flow comparable to the observations.

\section{\textbf{JETS AND  
   GRAVITATIONAL COLLAPSE}}

\begin{figure}[t!]
\plottwover{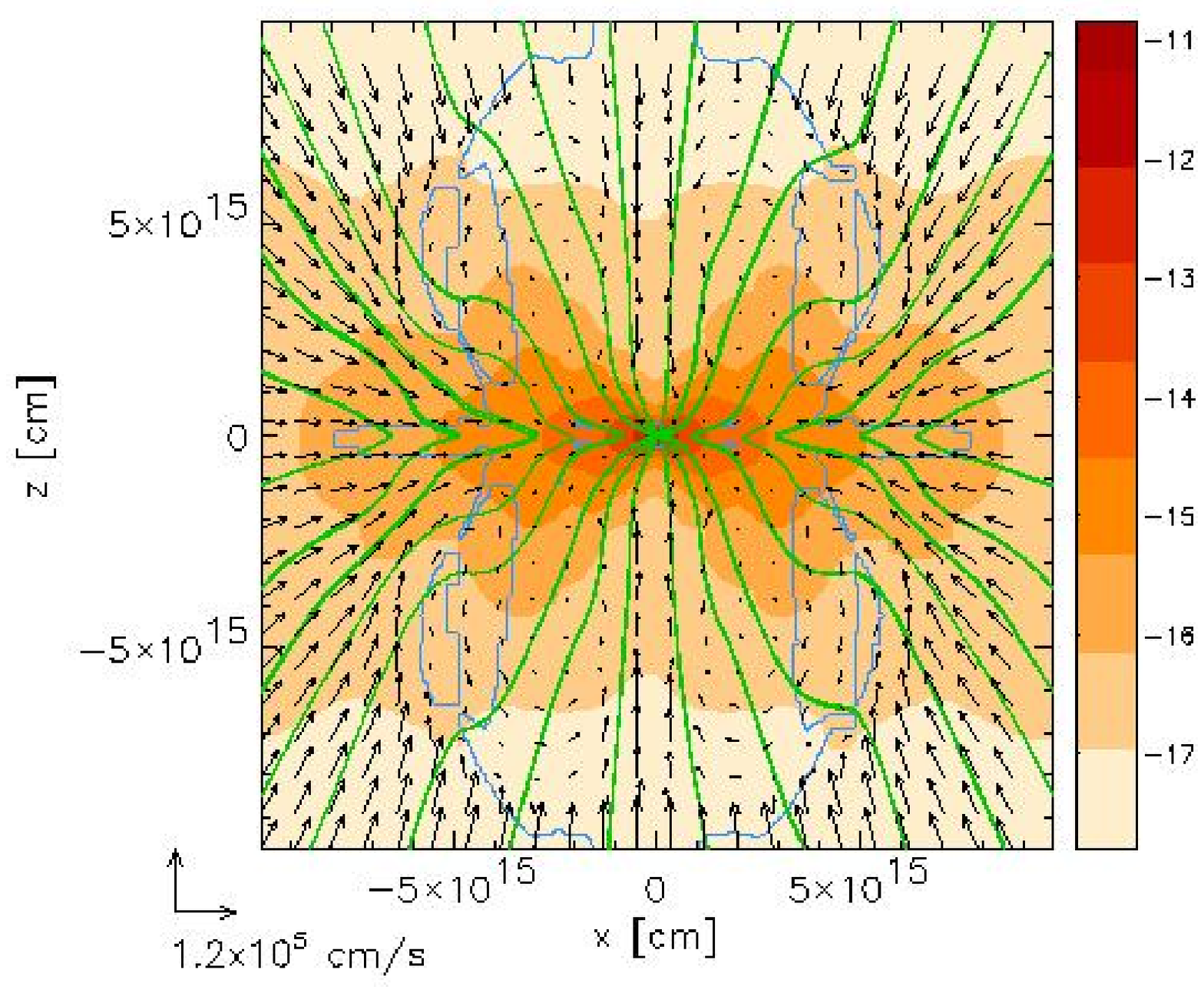}{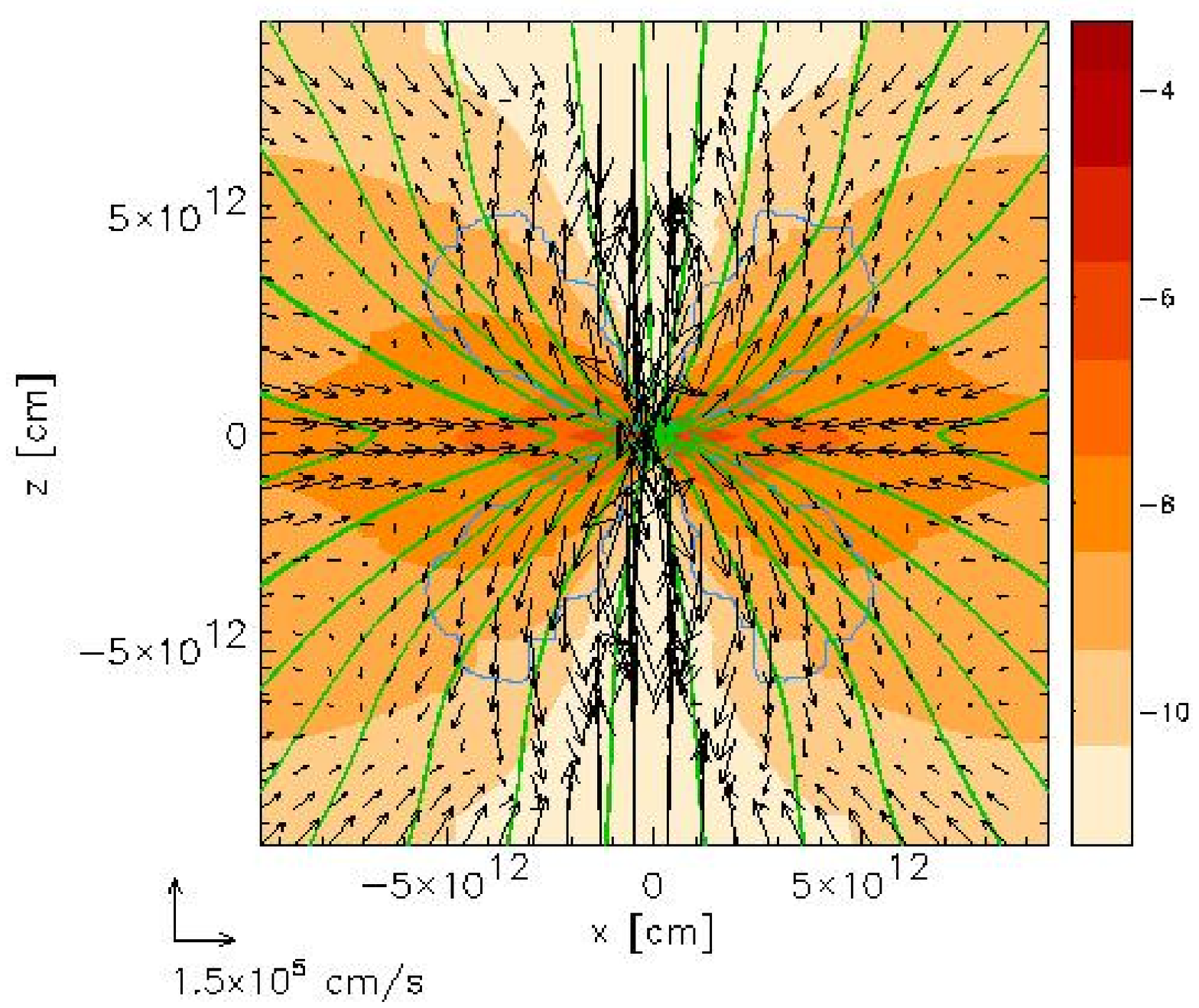}
\caption{Large scale outflow (top frame, scale of
hundreds of AU), and and small scale disk wind
and jet formed (bottom frame, scale of a fraction of 
an AU) during the gravitational collapse of a magnetized
B-E, rotating cloud core. Cross-sections through the disk
and outflows are shown -- the blue contour marks the Alf\'ven
surface. Snapshots taken of an Adaptive Mesh calculation
at about 70,000 years into the collapse.  [Adapted from {\it Banerjee
and Pudritz}, 2006].}
\label{fig:outflow}
\end{figure}

Jets are expected to be associated with gravitational collapse 
because disks are the result of the collapse of rotating
molecular cloud cores.
One of the first simulations to show
how jets arise during 
gravitational collapse 
is the work of {\it Tomisaka} (1998, 2002).
Here, the collapse of a  
magnetized core within a rotating cylinder of gas
gave rise the formation of a disk from which a centrifugally
drive, disk wind was produced. 
Analytical work by {\it Krasnopolsky and K\"onigl} (2002)
provides an interesting class of self-similar solutions for 
the collapse of rotating, magnetized, isothermal cores
including ambipolar diffusion that could be matched onto
BP82 outflows.  

Given the importance of Bonner-Ebert (B-E) spheres in the physics
of star formation and gravitational collapse, recent efforts
have focused on understanding the evolution of magnetized
B-E spheres.  Whereas purely hydrodynamic collapses of such
objects never show outflows (e.g., {\it Foster and Chevalier}, 1993;
{\it Banerjee et al.}, 2004), the addition of a magnetic field produces
them. {\it Matsumoto and Tomisaka} (2004) have studied the collapse
of rotating B-E spheres wherein the magnetic axis is inclined
with the initial rotation axis.  They observe that after the
formation of an adiabatic core, outflow is ejected along the
local magnetic field lines.  Eventually, strong torques result
in the rapid alignment of the rotation and magnetic field
vectors.  

The collapse of a magnetized, rotating, B-E sphere with 
molecular cooling included (but not dust), was carried out by 
{\it Banerjee and Pudritz} (2006) using the FLASH 
adaptive mesh refinement code.  The results of this
simulation are shown in Fig.~5 which shows the end state
(at about 70,000 yrs) of the collapse of a B-E sphere 
that is chosen to be precisely the Bok globule observed
by {\it Alves et al.}\ (2001) -- whose mass is $2.1 M_{\odot}$ and 
radius $R=1.25 \times 10^4$ AU at an initial temperature
of 16 K.  
Two types of outflow can be seen:
(i) an outflow that originates at scale of $
\simeq 130$ AU on the forming disk that consists of a 
wound up column of toroidal magnetic field whose pressure
gradient pushes out a slow outflow; and (ii) a disk-wind
that collimates into a jet on scale of 0.07 AU.  A 
tight proto-binary system has formed in this simulation,
whose masses are still very small $\le 10^{-2} M_{\odot}$,
which is much less than the mass of the disk at this
time $\simeq 10^{-1} M_{\odot}$.
The outer flow bears the hallmark of a magnetic tower, first
observed by {\it Uchida and Shibata}, and studied by 
{\it Lynden-Bell} (2003).  Both flow components are unbound,
with the disk wind 
reaching $3$ km s$^{-1}$ at 0.4 AU which is quite super-Alfv\'enic
and above the local escape speed.  
The outflow and jet speeds will 
increase as the central mass grows.  

We conclude that the theory and computation   
of jets and outflows is in excellent agreement  
with new observations of many kinds.  Disk winds are
triggered during magnetized collapse and persist throughout
the evolution of the disk.  They efficiently tap accretion power, 
transport a significant part portion of the disk's
angular momentum, and can achieve different degrees 
of collimation depending on their mass loading.  
Accretion-powered stellar winds may also
solve the stellar angular momentum problem.
We are optimistic that observations will
soon be able to test
the universality of outflows, all the way from circumplanetary disks
to those around O stars.

\textbf{ Acknowledgments.}  We are indebted to 
Nordita for 
hosting an authors' meeting,
Tom Ray for stimulating discussions, 
Robi Banerjee and Bo Reipurth 
for careful reads of the manuscript,   
and an anonymous referee for a very useful report.
The research of REP and RO is supported by grants from NSERC of Canada.
C.F. acknowledges travel support by the German science foundation
to participate in the PPV conference.

\centerline\textbf{REFERENCES}
\bigskip
\parskip=0pt
{\small
\baselineskip=11pt
 
\refs
Afshordi N., Mukhopadhyay B., and Narayan R. (2005)
{\it Astrophys. J., 629}, 373-382.

\refs
Alves J. F., Lada C. J., and Lada E. A. (2001) {\it Nature, 409}, 159-161.

\refs
Anderson J. M., Li Z.-Y., Krasnopolsky R., and Blandford R. D. (2003)
{\it Astrophys. J., 590}, L107-L110.

\refs
Anderson J. M., Li Z.-Y., Krasnopolsky R., and Blandford R. D. (2005)
{\it Astrophys. J., 630}, 945-957.

\refs
Bacciotti F., Mundt R., Ray T. P., Eisl\"offel J., Solf J.,
and Camenzind M. (2000) {\it Astrophys. J., 537}, L49-L52.

\refs
Bacciotti F., Ray T. P., Mundt R., Eisl\"offel J.,
and Solf J. (2002) {\it Astrophys. J., 576}, 222-231.

\refs
Bacciotti F., Ray T. P., Eisl\"offel J., Woitas J., Solf J.,
Mundt R., and Davis C. J. (2003) {\it Astrophys. Space Sci. 287}, 3-13.

\refs
Bacciotti F. (2004) {\it Astrophys. Space Sci. 293}, 37-44.

\refs
Balbus S. A., Hawley J. F., and Stone J. M. (1996)
{\it Astrophys. J., 467}, 76-86.

\refs
Banerjee R. and Pudritz R. E. (2006)
{\it Astrophys. J.}, 641, April 20.

\refs
Banerjee R., Pudritz R. E., and Holmes L. (2004)
{\it Mon. Not. R. Astron. Soc., 355}, 248-272.

\refs 
Bardou A., Rekowski B. v., Dobler W., Brandenburg A., and Shukurov A. (2001)
{\it Astron. Astrophys., 370}, 635-648.

\refs
Bateman G. (1980) {\it Magneto-Hydrodynamical Instabilities},
MIT Press, Cambridge.

\refs
Blandford R. D. and Payne D. G. (1982)
{\it Mon. Not. R. Astron. Soc., 199}, 883-903.

\refs
Boldyrev S. and Cattaneo F. (2004) {\it Phys. Rev. Lett., 92}, 144501.

\refs
Bourke T. L., Crapsi A., Myers P. C., Evans N. J., et al.\ (2005)
{\it Astrophys. J., 633}, L129-L132.

\refs
Brandenburg A. and Sokoloff D. (2002)
{\it Geophys. Astrophys. Fluid Dynam., 96}, 319-344.

\refs
Brandenburg A. and Subramanian K. (2005) {\it Phys. Rep., 417}, 1-209.

\refs
Brandenburg A., Nordlund \AA., Stein R. F., and Torkelsson, U. (1995)
{\it Astrophys. J., 446}, 741-754.

\refs
Cabrit S., Raga A. C., and Gueth F. (1997)
In {\it Herbig-Haro Flows and the Birth of Low Mass Stars}
(B. Reipurth and C. Bertout, eds.), pp.\ 163-180. Dordrecht, Kluwer.

\refs
Camenzind M. (1990)
{\em Reviews in Modern Astronomy, 3}, 234-265.

\refs
Campbell C. G. (2000) {\it Mon. Not. R. Astron. Soc., 317}, 501-527.

\refs
Campbell C. G. (2003) {\it Mon. Not. R. Astron. Soc., 345}, 123-143.

\refs
Campbell C. G., Papaloizou J. C. B., and Agapitou V. (1998)
{\it Mon. Not. R. Astron. Soc., 300}, 315-320.

\refs
Cao X. and Spruit H. C. 1994, {\it Astron. Astrophys., 287}, 80-86.

\refs
Casse F. and Ferreira J. (2000) {\it Astron. Astrophys., 353}, 1115-1128.

\refs
Casse F. and Keppens R. (2002) {\it Astrophys. J., 581}, 988-1001.

\refs
Casse F. and Keppens R. (2004) {\it Astrophys. J., 601}, 90-103.

\refs
Chagelishvili G. D., Zahn J.-P., Tevzadze A. G., and
Lominadze, J. G. (2003) {\it Astron. Astrophys., 402}, 401-407.

\refs
Chiang E. I., Joung M. K., Creech-Eakman M. J., Qi C., Kessler J. E.,
Blake G. A., and van Dishoeck E. F. (2001)
{\it Astrophys. J., 547}, 1077-1089.

\refs
Cowling T. G. (1934) {\it Mon. Not. R. Astron. Soc., 94}, 39-48.

\refs
De Villiers J.-P., Hawley J. F., Krolik J. H., and Hirose S. (2005),
{\it Astrophys. J., 620}, 878-888.

\refs
{Donati J.-F., Paletou F, Bouvier J., and Ferreira J. (2005),
{\it Nature, 438}, 466-469.

\refs
Dubrulle B., Marie L., Normand C., Richard D., Hersant F.,
and Zahn J.-P. (2005) {\it Astron. Astrophys., 429}, 1-13.

\refs
Eichler D. (1993), {\it Astrophys. J., 419}, 111-116.

\refs
Fendt C. (2005), {\it Astrophys. J.}, submitted (astro-ph/0511611). 

\refs
Fendt C. and Camenzind M. (1996), {\it Astron. Astrophys., 313}, 591-604.

\refs
Fendt C. (2003) {\it Astron. Astrophys., 411}, 623-635.

\refs
Fendt C. and Elstner D. (1999) {\it Astron. Astrophys., 349}, L61-L64.

\refs 
Fendt C. and Elstner D. (2000) {\it Astron. Astrophys., 363}, 208-222.

\refs
Fendt C. and {\v C}emelji{\' c}, M. (2002)
{\it Astron. Astrophys., 395}, 1045-1060.

\refs
Ferriera J. (1997) {\it Astron. Astrophys., 319}, 340-359.

\refs
Ferreira J. and Casse F. (2004) {\it Astrophys. J., 601}, L139-L142.

\refs
Fleming T. and Stone J. M. (2003) {\it Astrophys. J., 585}, 908-920.

\refs
Foster, P. N. and Chevalier, R. A. (1993)
{\it Astrophys. J., 416}, 303-311.

\refs
Fromang S., Terquem C., and Balbus S. A. (2002)
{\it Mon. Not. R. Astron. Soc., 329}, 18-28.

\refs
Gammie C. F. (1996) {\it Astrophys. J., 457}, 355-362.

\refs
Goodman J. and Rafikov R. R. (2001)
{\it Astrophys. J., 552}, 793-802.

\refs
Goodson A. P., Winglee R. M., and B\"ohm K-.H. (1997)
{\it Astrophys. J., 489}, 199-209.

\refs 
Goodson A. P., B\"ohm K.-H., and Winglee R. M. (1999)
{\it Astrophys. J., 524}, 142-158.

\refs 
Goodson A. P. and Winglee R. M. (1999)
{\it Astrophys. J., 524}, 159-168.

\refs
Greenhill L. J., Gwinn C. R., Schwartz C., Moran J. M., and
Diamond, P. J. (1998) {\it Nature, 396}, 650-653.

\refs
Hardee P. and Rosen A. (1999) {\it Astrophys. J., 524}, 650-666.

\refs
Hawley J. F., Balbus S. A., and Winters W. F. (1999)
{\it Astrophys. J., 518}, 394-404.

\refs
Hawley J. F. (2000) {\it Astrophys. J., 528}, 462-479.

\refs
Hayashi M. R., Shibata K., and Matsumoto R. (1996)
{\it Astrophys. J., 468}, L37-L40.

\refs
Herbst W., Bailer-Jones C. A. L., Mundt R., Meisenheimer K.,
and Wackermann R. (2002) {\it Astron. Astrophys., 398}, 513-532.

\refs
Heyvaerts J. (2003) In
{\it Accretion discs, jets, and high energy phenomena in
astrophysics}, (V. Beskin, et al., eds.), p.\ 3., Springer-Verlag, Berlin.

\refs
Heyvaerts J. and Norman C. (1989) {\it Astrophys. J., 347}, 1055-1081.

\refs
Inutsuka S.-I. and Sano T. (2005)
{\it Astrophys. J., 628}, L155-L158.

\refs
Keppens R. and Goedbloed J. P. (2000)
{\it Astrophys. J., 530}, 1036-1048.

\refs
Kigure H. and Shibata K. (2005)
{\it Astrophys. J., 634}, 879-900.

\refs
Kley W. and Lin D. N. C. (1992) {\it Astrophys. J., 397}, 600-612.

\refs
Kley W., D'Angelo G., and Henning T. (2001) {\it Astrophys. J., 547}, 457-464.

\refs 
K\"onigl A. (1991) {\it Astrophys. J., 370}, L39-L43.

\refs 
K\"onigl A. (1999) {\it New Astron. Rev., 43}, 67-77.

\refs
K\"onigl A. (2004) {\it Astrophys. J., 617}, 1267-1271.

\refs
K\"onigl A. and Choudhuri A. R. (1985) {\it Astrophys. J., 289}, 173-187.

\refs
K\"onigl A. and Pudritz R. E. (2000) In {\it Protostars and Planets IV}
(V. Mannings et al., eds.), p.\ 759 - 787. Univ. of Arizona, Tucson.

\refs
Krasnopolsky R. and K\"onigl A. (2002) {\it Astrophys. J., 580}, 987-1012.

\refs
Krasnopolsky, R., Li Z.-Y., and Blandford R. D. (1999)
{\it Astrophys. J., 526}, 631-642.

\refs
Krasnopolsky R., Li Z.-Y., and Blandford R. D. (2003)
{\it Astrophys. J., 595}, 631-642.

\refs
Krumholz M. R., McKee C. F., and Klein R. I. (2005)
{\it Astrophys. J., 618}, L33-L36.

\refs
Kudoh T., Matsumoto R., and Shibata K. (1998)
{\it Astrophys. J., 508}, 186-199.

\refs 
K\"uker M., Henning T., and R\"udiger G. (2003)
{\it Astrophys. J., 589}, 397-409; erratum: {\it Astrophys. J., 614}, 526.

\refs
Kuwabara T., Shibata K., Kudoh T., and Matsumoto, R. (2005)
{\it Astrophys. J., 621}, 921-931.

\refs
Lai D. (1999) {\it Astrophys. J., 524}, 1030-1047.

\refs
Lai D. (2003) {\it Astrophys. J., 591}, L119-L122.

\refs
Lee C.-F., Mundy L. G., Reipurth B., Ostriker E. C., and Stone, J. M. (2000)
{\it Astrophys. J., 542}, 925-945.

\refs
Lesur G. and Longaretti P.-Y. (2005) {\it Astron. Astrophys., 444}, 25-44.

\refs
Li Z.-Y. (1995) {\it Astrophys. J., 444}, 848-860.

\refs
Li Z.-Y. (1996) {\it Astrophys. J., 465}, 855-868.

\refs
Long M., Romanova M. M., and Lovelace, R. V. E. (2005)
{\it Astrophys. J., 634}, 1214-1222.

\refs
Lubow S. H., Papaloizou J. C. B., and Pringle J. E. (1994)
{\it Mon. Not. R. Astron. Soc., 268}, 1010-1014.

\refs
Lucek S. G. and Bell A. R. (1996)
{\it Mon. Not. R. Astron. Soc., 281}, 245-256.

\refs
Lynden-Bell D. (2003) {\it Mon. Not. R. Astron. Soc., 341}, 1360-1372.

\refs
Matsumoto R., Uchida Y., Hirose S., Shibata K., Hayashi M. R.,
Ferrari A., Bodo G., and Norman C. (1996) {\it Astrophys. J., 461}, 115-126. 

\refs
Matsumoto T. and Tomisaka K. (2004) {\it Astrophys. J., 616}, 266-282.

\refs
Matsumura S. and Pudritz R. E. (2005) {\it Astrophys. J., 618}, L137-L140.

\refs
Matsumura S. and Pudritz R. E. (2006)
{\it Mon. Not. R. Astron. Soc., 365}, 572-584.

\refs 
Matt S. and Pudritz R. E. (2005a) {\it Mon. Not. R. Astron. Soc., 356}, 167-182.

\refs
Matt S. and Pudritz, R. E. (2005b) {\it Astrophys. J., 632}, L135-L138.

\refs 
Matt S., Goodson A. P., Winglee R. M., and Bo\"hm K.-H. (2002)
{\it Astrophys. J., 574}, 232-245.

\refs
M\'enard F. and Duch\^ene G. (2004) {\it Astron. Astrophys., 425}, 973-980.

\refs
Meier D. L., Edgington S., Godon P., Payne D. G., and Lind, K. R. (1997)
{\it Nature, 388}, 350-352.

\refs
Michel F. C. (1969) {\it Astrophys. J., 158}, 727-738.

\refs 
Miller K. A. and Stone J. M. (1997) {\it Astrophys. J., 489}, 890-902.

\refs
Miller K. A. and Stone J. M. (2000) {\it Astrophys. J., 376}, 214-419.

\refs
Molemaker M. J., McWilliams J. C., and Yavneh I. (2001)
{\it Phys. Rev. Lett., 86}, 5270-5273.

\refs
Nakamura M., Uchida Y., and Hirose S. (2001)
{\it New Astron., 6}, 61-78.

\refs
Nakamura M. and Meier D. (2004) {\it Astrophys. J., 617}, 123-154.

\refs
Ouyed R. and Pudritz R. E. (1997a) {\it Astrophys. J., 482}, 712-732 (OPI).

\refs
Ouyed R. and Pudritz R. E. (1997b) {\it Astrophys. J., 484}, 794-809 (OPII).

\refs
Ouyed R., Pudritz R. E., and Stone, J. M. (1997)
{\it Nature, 385}, 409-414 (OPS).

\refs
Ouyed R. and Pudritz R. E. (1999)
{\it Mon. Not. R. Astron. Soc., 309}, 233-244 (OPIII).

\refs
Ouyed R., Clarke D. A., and Pudritz R. E. (2003)
{\it Astrophys. J., 582}, 292-319 (OCP).

\refs
Ouyed R. (2003) {\it Astrophys. Space Sci., 287}, 87-97.

\refs 
O'Sullivan M., Truss M., Walker C., Wood K., Matthews O., et al. (2005)
{\it Mon. Not. R. Astron. Soc., 358}, 632-640.

\refs
Pelletier G. and Pudritz R. E. (1992) {\it Astrophys. J., 394}, 117-138.

\refs
Pfeiffer H. P. and Lai D. (2004) {\it Astrophys. J., 604}, 766-774.

\refs
Pinte C. and M\'enard F. (2004)
In {\it The Search for Other Worlds} (S. S. Holt and D. Demings, eds.),
pp.\ 123-126. American Institute of Physics.

\refs
Pringle J. E. (1996) {\it Mon. Not. R. Astron. Soc. 281}, 357-361.

\refs
Pudritz R. E. (2003)
In {\it Accretion discs, jets, and high energy phenomena in 
astrophysics}, (V. Beskin, et al., eds.), p.\ 187 - 230. Springer-Verlag, Berlin.

\refs
Pudritz R. E. and Banerjee B. (2005)
In {\it Massive star birth: A crossroads of Astrophysics},
R. Cesaroni, M. Felli, E. Churchwell, and M. Walmsley eds., 
pp.163-173, Cambridge University, Cambridge.

\refs
Pudritz R. E., Rogers C., and Ouyed R. (2006)
{\it Mon. Not. R. Astron. Soc., 365}, 1131-1148.

\refs
Pudritz R. E. and Norman C. A. (1983) {\it Astrophys. J., 274}, 677-697.

\refs
Pudritz R. E. and Norman C. A. (1986) {\it Astrophys. J., 301}, 571-586.

\refs
Quillen A.C. and Trilling D. E. (1998) {\it Astrophys. J., 508}, 707-713.

\refs 
von Rekowski B., Brandenburg A., Dobler W., and Shukurov A. (2003)
{\it Astron. Astrophys., 398}, 825-844.

\refs 
von Rekowski B. and Brandenburg A. (2004) {\it Astron. Astrophys., 420}, 17-32.

\refs 
von Rekowski B. and Brandenburg A. (2006) {\it Astron. Nachr., 327}, 53-71.

\refs
Roberts P. H. (1967)
{\it Introduction to Magneto-Hydrodynamics}, Longmans, London.

\refs
Romanova M. M., Ustyugova G. V., Koldoba A. V., Chechetkin V. M.,
and Lovelace R. V. (1997) {\it Astrophys. J., 482}, 708-711.

\refs
Romanova M. M., Ustyugova G. V., Koldoba A. V., Chechetkin V. M.,
and Lovelace R. V. E. (1998)
{\it Astrophys. J., 500}, 703-713.

\refs
Romanova M. M., Ustyugova G. V., 
Koldoba A. V., and Lovelace R. V. E. (2002)
{\it Astrophys. J., 578}, 420-438.

\refs
Romanova M. M., Toropina O. D., 
Toropin Y. M., and Lovelace R. V. E. (2003)
{\it Astrophys. J., 588}, 400-407.

\refs 
Romanova M. M., Ustyugova G. V., 
Koldoba A. V., and Lovelace R. V. E. (2004)
{\it Astrophys. J., 610}, 920-932.

\refs
R\'o\.zyczka M. and Spruit H. C. 1993, {\it Astrophys. J., 417}, 677-686.

\refs
Sano T., Miyama S. M., Umebayashi T., and Nakano T.  (2000)
{\it Astrophys. J., 543}, 486-501.

\refs
Sano T. and Stone, J. M. (2002) {\it Astrophys. J., 570}, 314-328.

\refs
Schandl S. and Meyer F. (1994) {\it Astron. Astrophys., 289}, 149-161.

\refs
Schekochihin A. A., Haugen N. E. L., Brandenburg A., Cowley S. C.,
Maron J. L., and McWilliams J. C. (2005) {\it Astrophys. J., 625}, L115-L118.

\refs
Semenov D., Wiebe D., and Henning, T. (2004)
{\it Astron. Astrophys., 417}, 93-106.

\refs
Shakura N. I. and Sunyaev R. A. (1973) {\it Astron. Astrophys., 24}, 337-355

\refs
Shalybkov D. and R\"udiger G. (2005) {\it Astron. Astrophys., 438}, 411-417.

\refs
Shibata K. and Uchida Y. (1986) {\it Publ. Astron. Soc. Japan, 38}, 631-660.

\refs 
Shu F. H., Najita J. R., Shang H., and Li Z.-Y. (2000)
In {\it Protostars and Planets IV}
(V. Mannings et al., eds.), p.\ 789 - 813. Univ. of Arizona, Tucson.

\refs 
Shu F., Najita J., Ostriker E., 
Wilkin F,. Ruden S., and Lizano S. (1994)
{\it Astrophys. J., 429}, 781-796.

\refs
Spruit H. C., Foglizzo T., and Stehle R. (1997)
{\it Mon. Not. R. Astron. Soc., 288}, 333-342.

\refs
Spruit H. C. and Uzdensky D. A. (2005) {\it Astrophys. J., 629}, 960-968.

\refs
Stone J. M., Hawley J. F., Gammie C. F., and Balbus S. A. (1996)
{\it Astrophys. J., 463}, 656-673.

\refs
Stone J. M. and Norman M. L. (1992) {\it Astrophys. J. Suppl., 80}, 753-790.

\refs
Stone J. M. and Norman M. L. (1994) {\it Astrophys. J., 433}, 746-756.

\refs 
Symington N. H., Harries T. J., and Kurosawa R. (2005a)
{\it Mon. Not. R. Astron. Soc., 356}, 1489-1500.

\refs 
Symington N. H., Harries T. J., Kurosawa R., and Naylor T. (2005b)
{\it Mon. Not. R. Astron. Soc., 358}, 977-984.

\refs
Todo Y., Uchida Y., Sato T., and Rosner R. (1993)
{\it Astrophys. J., 403}, 164-174.

\refs
Tomisaka K. (1998) {\it Astrophys. J., 502}, L163-L167.

\refs
Tomisaka K. (2002) {\it Astrophys. J., 575}, 306-326.

\refs
Turner N. J. (2004) {\it Astrophys. J., 605}, L45-L48.

\refs 
Uchida Y. and Shibata K. (1984)
{\em Publ. Astron. Soc. Japan, 36}, 105-118.

\refs
Uchida Y. and Shibata K. (1985)
{\it Proc. Astron. Soc. Japan, 37}, 515-535.

\refs
Uchida Y., McAllister A., Strong K. T., Ogawara Y., Shimizu T.,
Matsumoto R., and Hudson H. S. (1992)
{\it Proc. Astron. Soc. Japan, 44}, L155-L160.

\refs
Umurhan O. M. (2006) {\it Mon. Not. R. Astron. Soc., 365}, 85-100.

\refs
Uzdensky D. A., K\"onigl A., and Litwin C. (2002)
{\it Astrophys. J., 565}, 1191-1204.

\refs
Vitorino B. F., Jatenco-Pereira V., and Opher R. (2002)
{\it Astron. Astrophys., 384}, 329.
 
\refs
Wardle M. and K\"onigl A. (1993) {\it Astrophys. J., 410}, 218-238.

\refs
Woitas J., Ray T. P., Bacciotti F.,  Davis C. J., and Eisl\"offel J. (2002)
{\it Astrophys. J., 580}, 336-342.

\refs
Wu Y., Wei Y., Zhao M., Shi Y., Yu W., Qin S., and Huang M. (2004)
{\it Astron. Astrophys., 426}, 503-515.

\refs
Ziegler U. and R\"udiger G. (2000) {\it Astron. Astrophys., 356}, 1141-1148.

\end{document}